\documentclass[twoside,11pt]{article}
\usepackage{jair, theapa, rawfonts}
\usepackage{times}
\usepackage{amsmath}
\usepackage{amsfonts}
\usepackage{amssymb}
\usepackage{verbatim}
\usepackage{thanks}
\usepackage{latexsym}

\newcommand{\bfe}[1]{\begin{bfseries}\emph{#1}\end{bfseries}\index{#1}}

\newcommand{\myra}{\mbox{$\:\rightarrow\:$}}

\newcommand{\fa}{\mbox{$\forall$}}

\newcommand{\LL}{\mbox{$\ldots$}}

\newcommand{\C}[1]{\mbox{$\{{#1}\}$}}           

\newcommand{\NI}{\noindent}

\newcommand{\II}{\vspace{2 mm}}

\newcommand{\szkew}[1]{\relax \setbox0=\hbox{\kern -24pt $\displaystyle#1$\kern 0pt }%
\box0}
{\catcode`\@=11 \global\let\ifjusthvtest@=\iffalse}

\newcounter{oldmycaption}

\def\smallromani{\renewcommand{\theenumi}{\roman{enumi}}
\renewcommand{\labelenumi}{(\theenumi)}}

\newtheorem{theorem}{Theorem}[section]
\newtheorem{defined}[theorem]{Definition}
\newenvironment{definition}{\begin{defined} \rm}{\end{defined}}
\newtheorem{exa}[theorem]{Example}
\newenvironment{example}{\begin{exa} \rm}{\end{exa}}
\newtheorem{proposition}[theorem]{Proposition}

\newtheorem{lemma}[theorem]{Lemma}
\newtheorem{corollary}[theorem]{Corollary}

\newtheorem{exe}{Exercise}

\newtheorem{pro}{Problem}

\newcounter{symbol}
\newcommand{\indexsyma}[1]%
{\stepcounter{symbol}\index{zzz1 \thesymbol @\protect#1}}
\newcommand{\indexsymb}[1]%
{\stepcounter{symbol}\index{zzz2 \thesymbol @\protect#1}}
\newcommand{\indexsymc}[1]%
{\stepcounter{symbol}\index{zzz3 \thesymbol @\protect#1}}
\newcommand{\indexsymd}[1]%
{\stepcounter{symbol}\index{zzz4 \thesymbol @\protect#1}}
\newcommand{\indexsyme}[1]%
{\stepcounter{symbol}\index{zzz5 \thesymbol @\protect#1}}

\newcommand{\name}[1]{\index{zz #1@#1}}

\newcommand{\eat}[1]{}

\renewenvironment{Proof}
      {\medskip\noindent{\bf Proof.}}
      {\hfill$\Box$\medskip}
\def\smallromani{\renewcommand{\theenumi}{\roman{enumi}}
\renewcommand{\labelenumi}{(\theenumi)}}

\jairheading{?}{????}{?-??}{?/??}{?/??}
\ShortHeadings{Undominated Groves Mechanisms}
{Guo, Markakis, Apt, Conitzer}
\firstpageno{1}

\begin{document}

\author{\name Mingyu Guo\email Mingyu.Guo@liverpool.ac.uk\\
\addr University of Liverpool, UK\\
\name Evangelos Markakis\email markakis@gmail.com\\
\addr Athens University of Economics and Business, Greece\\
\name Krzysztof R. Apt\email apt@cwi.nl\\
\addr CWI and University of Amsterdam, the Netherlands\\
\name Vincent Conitzer \email conitzer@cs.duke.edu\\
\addr Duke University, USA
}



\title{Undominated Groves Mechanisms}

\date{}
\maketitle

\begin{abstract} 
The family of Groves mechanisms, which includes the well-known
VCG mechanism (also known as the Clarke mechanism), is a family of efficient
and strategy-proof mechanisms.  Unfortunately, the Groves mechanisms are
generally not budget balanced. That is, under such mechanisms, payments may
flow into or out of the system of the agents, resulting in deficits or reduced
utilities for the agents.  We consider the following problem: within the family
of Groves mechanisms, we want to identify mechanisms that give the agents the
highest utilities, under the constraint that these mechanisms must never incur
deficits.
\let\thefootnote\relax\footnotetext{Some of this work appeared in preliminary form in \cite{Apt08:Welfare} and \cite{Guo08:Undominated}.}

We adopt a prior-free approach. We introduce two general measures for comparing
mechanisms in prior-free settings. We say that a non-deficit Groves mechanism
$M$ {\em individually dominates} another non-deficit Groves mechanism $M'$ if
for every type profile, every agent's utility under $M$ is no less than that
under $M'$, and this holds with strict inequality for at least one type profile
and one agent.  We say that a non-deficit Groves mechanism $M$ {\em
collectively dominates} another non-deficit Groves mechanism $M'$ if for every
type profile, the agents' total utility under $M$ is no less
than that under $M'$, and this holds with strict inequality for at least one
type profile.  The above definitions induce two partial orders on non-deficit
Groves mechanisms.  We study the maximal elements corresponding to these two
partial orders, which we call the {\em individually undominated} mechanisms and
the {\em collectively undominated} mechanisms, respectively.  \end{abstract}

\section{Introduction}
\label{sec:intro}


Mechanism design is often employed for coordinating group decision making among
agents.  Often, such mechanisms impose payments that agents have to pay to a
central authority. Although maximizing revenue is a desirable objective in many
settings (for example, if the mechanism is an auction designed by the seller),
it is not desirable in situations where no entity is profiting from the
payments. Some examples include public project problems as well as certain
resource allocation problems without a seller ({\em e.g.}, the right to use a
shared good in a given time slot, or the assignment of take-off slots among
airline companies). In such cases, we would like to have mechanisms that
minimize payments (or, even better, achieve budget balance), while maintaining
other desirable properties, such as being efficient, strategy-proof and
non-deficit ({\em i.e.}, the mechanism does not need to be funded by an
external source).

The family of Groves mechanisms, which includes the well-known VCG mechanism
(also known as the Clarke mechanism), is a family of efficient and
strategy-proof mechanisms.  In many sufficiently general settings, including the settings
that we will study in this paper, the Groves mechanisms are the only efficient
and strategy-proof mechanisms~\cite{Holmstrom79:Groves}.  Unfortunately though,
the Groves mechanisms are generally not budget balanced. That is, under such
mechanisms, payments may flow into or out of the system of the agents,
resulting in deficits or reduced utilities for the agents.  Motivated by this we consider 
in this paper the following problem: within the family of Groves mechanisms,
we want
to identify mechanisms that give the agents the highest utilities, under the
constraint that these mechanisms never incur deficits.\footnote{The
agents' utilities may be further increased if we also consider mechanisms
outside of the Groves
family~\cite{Guo08:Better,Clippel09:Destroy,Faltings05:Budget,Guo11:Budget}, but in this
paper we take efficiency as a hard constraint.}

We adopt a prior-free approach, where each agent $i$ knows only his own valuation $v_i$, and there is no prior probability 
distribution over the other agents' values.
We introduce two natural measures for comparing
mechanisms in prior-free settings. Given a performance indicator, we say that
mechanism $M$ {\em individually dominates} mechanism $M'$ if for every type
profile of the agents, $M$ performs no worse than $M'$ from the perspective of each individual agent,
and this holds with strict inequality for at least one type profile and one
agent. We say that mechanism $M$ {\em collectively dominates} mechanism $M'$ if
for every type profile, $M$ performs no worse than $M'$ from the perspective of
the set of agents as a whole, and this holds with strict inequality for at least one
type profile.  In this paper, we focus on maximizing the agents' utilities.
Given this specific performance indicator, individual and collective dominance are determined 
by comparing either individual utilities or the sum of the agents' utilities, respectively.

The above definitions induce two partial orders on non-deficit Groves
mechanisms.  Our goal in this work is to identify and study the maximal
elements corresponding to these two partial orders, which we call the {\em
individually undominated} (non-deficit Groves) mechanisms and the {\em
collectively undominated} (non-deficit Groves) mechanisms, respectively.  It
should be noted that the partial orders we focus on may be different from the
partial orders induced by other performance indicators, {\em e.g.}, if the criterion
is the revenue extracted from the agents. 


\subsection{Structure of the paper}

The presentation of our results is structured as follows: In
Sections~\ref{sec:prelim} and~\ref{sec:groves}, we formally define the notions
of individual and collective dominance, as well as the family of Groves mechanisms, and
then provide some basic observations.  We also establish some general
properties of anonymous Groves mechanisms which we use later on, and which may
be of independent interest.  We then begin our study of individual dominance in
Section~\ref{sec:individual}, where we give a characterization of individually
undominated mechanisms. We also propose two techniques for transforming any
given non-deficit Groves mechanism into one that is individually undominated. 

In Sections~\ref{sec:multi-unit} and~\ref{sec:pp-equal} 
we study the question of finding collectively
undominated mechanisms in two settings. The first (Section~\ref{sec:multi-unit}) is auctions of multiple
identical units with unit-demand bidders.  In this setting,
the VCG mechanism is collectively dominated by other non-deficit Groves mechanisms, such
as the Bailey-Cavallo mechanism~\cite{Bailey97:Demand,Cavallo06:Optimal}.
We obtain a complete characterization of collectively undominated mechanisms that
are anonymous and linear (meaning that the redistribution is a linear function of the ordered type profile; see Section~\ref{sec:multi-unit}
for the definition). In particular, we show that the collectively undominated mechanisms that are anonymous and linear are exactly the {\em
Optimal-in-Expectation Linear (OEL) redistribution mechanisms}, which include
the Bailey-Cavallo mechanism and were introduced in~\cite{Guo10:Optimal}.
The second setting (Section~\ref{sec:pp-equal}) is public project problems, where a set of agents must
decide on whether and how to finance a project ({\em e.g.}, building a bridge).  We show that in
the case where the agents have identical participation costs, the VCG mechanism is collectively undominated. 
On the other hand, when the
participation costs can be different across agents, there exist mechanisms that
collectively dominate VCG.
We finally show that when the participation costs
are different across agents, the VCG mechanism remains collectively undominated
among all pay-only mechanisms.

\subsection{Related work}
\label{subsec:relwork}

How to efficiently allocate resources among a group of competing agents is a
well-studied topic in economics literature. For example, the famous
Myerson-Satterthwaite Theorem~\cite{Myerson83:Efficient} rules out the
existence of efficient, Bayes-Nash incentive compatible, budget-balanced, and
individually rational mechanisms.  Cramton et al.~\citeyear{Cramton87:Dissolving} characterized the
Bayes-Nash incentive compatible and individually rational mechanisms for
dissolving a partnership, and gave the necessary and sufficient condition for
the possibility of dissolving partnership
efficiently. 

The main difference between these papers and ours is that we adopt a
prior-free approach. That is, we do not assume that we know the prior
distribution of the agents' valuations. As a result of this, our
notion of truthfulness is strategy-proofness, which is stronger than
Bayes-Nash incentive compatibility.  In many sufficiently general
settings, including the settings that we will study in this paper, the
Groves mechanisms are the only efficient and strategy-proof
mechanisms~\cite{Holmstrom79:Groves}.  That is, the search of
undominated Groves mechanisms is, in many settings, the search of
efficient, strategy-proof, and non-deficit mechanisms that are closest
to budget-balance.

Recently, there has been a series of works on VCG redistribution mechanisms,
which are mechanisms that make social decisions according to the efficient and
strategy-proof VCG mechanism, and then redistribute some of the VCG payments
back to the agents, under certain constraints, such as that an agent's
redistribution should be independent of his own type (therefore ensuring
strategy-proofness), and that the total redistribution should never exceed the
total VCG payment (therefore ensuring non-deficit). Actually, any non-deficit
Groves mechanism can be interpreted as such a VCG-based redistribution mechanism, and any
(non-deficit) VCG redistribution mechanism corresponds to a non-deficit Groves
mechanism (more details on this are provided in Section~\ref{sec:prelim}).

One example of a redistribution mechanism is the Bailey-Cavallo (BC)
mechanism~\cite{Cavallo06:Optimal}.\footnote{In settings that are revenue
monotonic, the Cavallo mechanism~\cite{Cavallo06:Optimal} coincides with 
a mechanism discovered earlier by Bailey~\citeyear{Bailey97:Demand}. The Bailey-Cavallo mechanism for a single-item auction was also
independently discovered in~\cite{Porter04:Fair}.} Under the BC mechanism,
every agent, besides participating in the VCG mechanism, also receives $\frac1n$
times the minimal VCG revenue that could have been obtained by changing this agent's
own bid.
In some settings ({\em e.g.}, a single-item auction), the BC mechanism can
successfully redistribute a large portion of the VCG payments back to the
agents. That is, in such settings, the BC mechanism both individually and
collectively dominates the VCG mechanism.  

Guo and Conitzer~\cite{Guo07:Worst} proposed another VCG redistribution
mechanism called the worst-case optimal (WCO) redistribution mechanism, in the
setting of multi-unit auctions with nonincreasing marginal values. WCO is
optimal in terms of the fraction of total VCG payment redistributed in the
worst case.\footnote{This notion of worst-case optimality was also studied
for more general settings in~\cite{Gujar11:Redistribution,Guo11:VCG,Guo12:Worst}.} Moulin~\cite{Moulin09:Almost}
independently derived WCO under a slightly different worst-case optimality
notion (in the more restrictive setting of multi-unit auctions with unit demand
only).  Guo and Conitzer~\cite{Guo10:Optimal} also proposed a family of VCG
redistribution mechanisms that aim to maximize the {\em expected} amount of VCG
payment redistributed, in the setting of multi-unit auctions with unit demand.
The members of this family are called the Optimal-in-Expectation Linear (OEL) redistribution
mechanisms. 

Finally, the paper that is the closest to what we study here is an early work
by Moulin on collectively undominated non-deficit Groves
mechanisms~\cite{Moulin86:Characterization}. It deals with the problem of
selecting an efficient public decision out of finitely many {\bf costless} alternatives.\footnote{In our public
project model, there is a cost associated with building the project.}
  Each
agent submits to the central authority his utility for each alternative.
Subsequently, the central authority makes a decision that maximizes the social
welfare.  Moulin showed \cite<Lemma 2 of>{Moulin86:Characterization} that the
VCG mechanism is collectively undominated in the above setting. This result
generalizes an earlier result for the case of two public decisions by Laffont
and Maskin~\cite{Laffont82:The}.

\section{Preliminaries}\label{sec:prelim}

We first briefly review payment-based mechanisms~\cite<see, {\em e.g.},>{Mas-Colell95}.

\subsection{Payment-based mechanisms}

Assume that there is a set of possible outcomes or \bfe{decisions} $D$,
a set $\{1, \LL, n\}$ of agents where $n \geq 2$, and for each agent $i$,
a set of \bfe{types} $\Theta_i$ and an (\bfe{initial})
\bfe{utility function} $ v_i : D \times \Theta_i \myra \mathbb{R}$. Let
$\Theta := \Theta_1 \times \cdots \times \Theta_n$.

In a (direct revelation) mechanism, each agent reports a type
$\theta_i\in\Theta_i$ and based on this, the mechanism selects an outcome and a
payment to be made by every agent. Hence a mechanism is given by a
pair of functions $(f,t)$, where $f$ is the decision function and $t$ 
is the payment function that determines the agents'
payments, {\em i.e.}, $f: \Theta \myra D$, and $t: \Theta \myra
\mathbb{R}^n$.

We put $t_i(\theta) := (t(\theta))_i$, {\em i.e.}, the function $t_i$ computes the payment of agent $i$.
%
For each vector $\theta$ of announced types, if
$t_i(\theta) \geq 0$, agent $i$ \bfe{pays} $t_i(\theta)$, and
if $t_i(\theta) < 0$, he \bfe{receives} $|t_i(\theta)|$.
When the true type of agent $i$ is $\theta_i$ and his
announced type is $\theta'_i$, his \bfe{final utility function} is 
defined by 
\[
u_i((f,t)(\theta'_i,
\theta_{-i}), \theta_i) := v_i(f(\theta'_i, \theta_{-i}), \theta_i) -
t_i(\theta'_i, \theta_{-i}),
\] 
where $\theta_{-i}$ is the vector of types
announced by the other agents.

\subsection{Properties of payment-based mechanisms}

We say that a payment-based mechanism $(f,t)$ is
  \begin{enumerate}
  \item[$\bullet$] \bfe{efficient} if for all $\theta \in
\Theta$ and $d \in D$, 
$\sum_{i = 1}^{n} v_i(f(\theta), \theta_i) \geq \sum_{i = 1}^{n} v_i(d, \theta_i)$, 
  \item[$\bullet$] \bfe{budget-balanced} if $\sum_{i = 1}^{n} t_i(\theta) = 0$ for all $\theta\in\Theta$,

  \item[$\bullet$] \bfe{non-deficit} if $\sum_{i = 1}^{n} t_i(\theta) \geq 0$ for all $\theta$, {\em i.e.}, the mechanism does not need to be funded by an external source,

  \item[$\bullet$] \bfe{pay-only} if $t_i(\theta) \geq 0$ for all $\theta$ and all $i \in \{1, \LL, n\}$,
  
  \item[$\bullet$] \bfe{strategy-proof} 
if for all $\theta$, $i \in \C{1,\LL,n}$, and
$\theta'_i$,
\[
u_i((f,t)(\theta_i, \theta_{-i}), \theta_i) \geq
u_i((f,t)(\theta'_i, \theta_{-i}), \theta_i),
\]
{\em i.e.}, for each agent $i$, reporting a false type, here $\theta'_i$, is not profitable.
  \end{enumerate}

\subsection{Individual and collective dominance}


We consider prior-free settings, where each agent $i$ knows only his own function $v_i$, and there is no belief or prior probability 
distribution regarding the other agents' initial utilities. Payment-based mechanisms 
can naturally be compared in terms of either the effect on each individual agent or the global effect on the whole set of agents.
We therefore introduce two measures for
comparing such mechanisms. Given a performance
indicator\footnote{By a performance indicator we mean a function of the mechanism's outcome that serves as a measure for comparing mechanisms. {\em E.g.}, it can be the final utility of an agent, or an arbitrary function of it, or a function of the agent's payment or any other function that depends on the decision rule and the payment rule of the mechanism.}, we say that mechanism $(f',t')$ {\em individually dominates}
mechanism $(f,t)$ if for every type profile, $(f',t')$ performs no worse than
$(f,t)$ from the perspective of every agent, and this holds with strict
inequality for at least one type profile and one agent. We say that mechanism
$(f',t')$ {\em collectively dominates} mechanism $(f,t)$ if for every type
profile, $(f',t')$ performs no worse than $(f,t)$ from the perspective of the
whole agent system, and this holds with strict inequality for at least one type
profile.  In this paper, we focus on maximizing the agents' utilities.  Given
this specific performance indicator, individual and collective dominance are captured by
the following definitions:

\begin{definition}
Given two payment-based mechanisms $(f,t)$ and $(f',t')$, we say
that $(f',t')$ \bfe{individually dominates} $(f,t)$ 
if 
\begin{itemize}
\item[$\bullet$] for all $\theta \in \Theta$ and all $i \in \{1, \LL, n\}$,
$u_i((f,t)(\theta), \theta_i) \leq  u_i((f',t')(\theta), \theta_i),
$
\item[$\bullet$] for some $\theta \in \Theta$ and some $i \in \{1, \LL, n\}$,
$u_i((f,t)(\theta), \theta_i) <  u_i((f',t')(\theta), \theta_i).$
\end{itemize} 
\end{definition}



\begin{definition}
Given two payment-based mechanisms $(f,t)$ and $(f',t')$, we say
that $(f',t')$ \bfe{collectively dominates} $(f,t)$ 
if
\begin{itemize}
\item[$\bullet$] for all $\theta \in \Theta$,
$
\sum_{i = 1}^{n} u_i((f,t)(\theta), \theta_i) \leq \sum_{i = 1}^{n} u_i((f',t')(\theta), \theta_i),
$

\item[$\bullet$] for some $\theta \in \Theta$,
$
\sum_{i = 1}^{n} u_i((f,t)(\theta), \theta_i) < \sum_{i = 1}^{n} u_i((f',t')(\theta), \theta_i).
$
\end{itemize}
\end{definition}

For two payment-based mechanisms $(f,t)$ and $(f',t')$, clearly if $(f',t')$
individually dominates $(f,t)$, then it also collectively dominates $(f,t)$.
Theorem~\ref{thm:diff} shows that the reverse implication however does not need
to hold, even if we limit ourselves to special types of mechanisms.  That is,
the fact that $(f',t')$ collectively dominates $(f,t)$ does not imply that
$(f',t')$ individually dominates $(f,t)$.  

Given a set $Z$ of payment-based mechanisms, individual and collective dominance induce two partial orders on 
$Z$, and we are interested in studying the maximal elements with respect to these partial orders. 
A maximal element with respect to the first partial order will be called an {\em individually undominated} mechanism,
{\em i.e.}, it is a mechanism that is not individually dominated by any other mechanism in $Z$.
A maximal element for the second partial order will be called a {\em collectively undominated} mechanism, {\em i.e.}, it is a mechanism that is not
collectively dominated by any other mechanism in $Z$. 
The maximal elements with respect to the two partial orders may differ and in particular, the notion of collectively undominated mechanisms
is generally a stronger notion. Clearly,
if $(f',t')\in Z$ is collectively undominated, then it is also individually undominated.   
The reverse may not be true, examples of which are provided in
Section~\ref{subsec:distinct}.

If we focus on the same decision function $f$, then individual and collective
dominance are strictly due to the difference of the payment functions.
Hence, $(f,t')$ individually dominates $(f,t)$ (or simply $t'$
individually dominates $t$) if and only if

\begin{itemize}
\item[$\bullet$] for all $\theta \in \Theta$ and all $i \in \{1, \LL, n\}$,
$t_i(\theta) \geq  t'_i(\theta)$, and
\item[$\bullet$] for some $\theta \in \Theta$  and some $i \in \{1, \LL, n\}$,
$t_i(\theta) > t'_i(\theta)$,
\end{itemize}
and $t'$ collectively dominates $t$ if 

\begin{itemize}
\item[$\bullet$] for all $\theta \in \Theta$, $\sum_{i =
  1}^{n} t_i(\theta) \geq \sum_{i = 1}^{n} t'_i(\theta)$, and

\item[$\bullet$] for some $\theta \in \Theta$, $\sum_{i =
  1}^{n} t_i(\theta) > \sum_{i = 1}^{n} t'_i(\theta)$.
\end{itemize}

We now define two transformations on payment-based mechanisms originating from
the same decision function.  Both transformations build upon the
surplus-guarantee concept \cite{Cavallo06:Optimal} for the specific case of the
VCG mechanism\footnote{The first transformation was originally defined in
\cite{Bailey97:Demand} and \cite{Cavallo06:Optimal} for the specific case
of the VCG mechanism and in \cite{Guo08:Undominated} for non-deficit Groves mechanisms.
We call it the \bfe{BCGC transformation} after the authors of these
papers (Bailey, Cavallo, Guo, Conitzer).}.


Consider a payment-based mechanism $(f,t)$. Given $\theta =
(\theta_1,\LL,\theta_n)$, let $T(\theta)$ be the total amount of payments,
{\em i.e.}, $T(\theta) := \sum_{i=1}^{n} t_i(\theta)$.  For each $i \in
\{1,\LL,n\}$ let
\[
S^{BCGC}_{i}(\theta_{-i}) := \inf_{\theta_i' \in \Theta_i} T(\theta'_i, \theta_{-i}).
\]
In other words, $S^{BCGC}_{i}(\theta_{-i})$ is the surplus guarantee independent of the report of agent $i$.
We then define the payment-based mechanism
$(f, t^{BCGC})$ by setting for $i \in \{1, \LL, n\}$
\[
t_i^{BCGC}(\theta) := t_i(\theta) - \frac{S^{BCGC}_i(\theta_{-i})}{n}.
\]

Also, for a fixed agent $j$, we define the payment-based mechanism
$(f, t^{BCGC(j)})$ by setting for $i\in \{1,\LL,n\}$
\[
t_i^{BCGC(j)}(\theta) := 
\left\{ 
  \begin{array}{l l}
     t_i(\theta) - S^{BCGC}_i(\theta_{-i})
     & \quad \text{if $i=j$}\\
     t_i(\theta) & \quad \text{if $i\neq j$}\\
  \end{array} \right.
\]

After the first transformation (from $(f,t)$ to $(f,t^{BCGC})$), every agent 
receives an additional\footnote{Receiving an additional positive amount means paying 
less and receiving an additional negative amount means paying more.} amount of 
$\frac1n$ times the surplus guarantee independent
of his own type.  During the second transformation (from $(f,t)$ to
$(f,t^{BCGC(j)}$), agent $j$ is chosen to be the only agent who receives an
additional amount. This additional amount equals the entirety of
the surplus guarantee independent of $j$'s own type. For both
transformations the agents' additional payments are independent of their own
types, thus the strategy-proofness is maintained: if $(f,t)$ is
strategy-proof, then so are $(f,t^{BCGC})$ and $(f,t^{BCGC(j)})$ for all $j$.


The following observations generalize some of the results of
\cite{Bailey97:Demand,Cavallo06:Optimal}.

\begin{proposition} \label{prop:bc}
\mbox{} \\[-6mm]
\begin{enumerate} \smallromani
\item Each payment-based mechanism of the form $t^{BCGC}$ is non-deficit.  

\item If $t$ is non-deficit, then either $t$ and $t^{BCGC}$ coincide or  $t^{BCGC}$ individually (and hence also collectively) dominates $t$.
\end{enumerate}

\end{proposition}

\begin{Proof}
$(i)$
For all $\theta$ and $i \in \{1, \LL, n\}$ we have $T(\theta) \geq S^{BCGC}_{i}(\theta_{-i})$, so
\begin{eqnarray*}
T^{BCGC}(\theta) & = &\sum_{i=1}^{n} t_i^{BCGC}(\theta) = T(\theta) - \sum_{i=1}^{n} \frac{S^{BCGC}_i(\theta_{-i})}{n} \\
& = &
\sum_{i=1}^{n} \frac{T(\theta) - S^{BCGC}_i(\theta_{-i})}{n} \geq 0.
\end{eqnarray*}

\NI
$(ii)$ If $t$ is non-deficit, then for all $\theta$ and all $i \in \{1, \LL, n\}$
we have $S^{BCGC}_i(\theta_{-i}) \geq 0$, and hence
$t_i^{BCGC}(\theta) \leq t_i(\theta)$.
\end{Proof}

The same claims hold for $t^{BCGC(j)}$ for $j \in \{1, \LL, n\}$, with equally simple proofs.



\section{Groves mechanisms}
\label{sec:groves}

We first briefly review Groves mechanisms.

\subsection{Preliminaries}

Recall that a \bfe{Groves mechanism}~\cite{Groves73:Incentives}
is a payment-based mechanism $(f,t)$ such that the following
hold\footnote{Here and below $\sum_{j\not=i}$ is a shorthand for the
  summation over all $j \in \{1,\LL,n\}, \ j \not=i$.}:

\begin{itemize}

\item[$\bullet$] $f(\theta)\in \arg\max_{d\in D} \sum_{i=1}^n v_i(d,\theta_i)$, 
i.e., the chosen outcome maximizes the \bfe{allocation welfare} (the agents' total valuation),

\item[$\bullet$] $t_i : \Theta \myra \mathbb{R}$ is defined by $t_i(\theta) :=  h_i(\theta_{-i})- g_i(\theta)$, where

\item[$\bullet$] $g_i(\theta) := \sum_{j \neq i} v_j(f(\theta), \theta_j)$,
  
\item[$\bullet$] $h_i: \Theta_{-i} \myra \mathbb{R}$ is an arbitrary function.

\end{itemize}




So $g_i(\theta)$ represents the allocation welfare from the
decision $f(\theta)$ with agent $i$ ignored.
Recall now the following crucial result~\cite<see, {\em e.g.},>{Mas-Colell95}.  
\II

\NI \textbf{Groves Theorem \cite{Groves73:Incentives}} Every Groves mechanism is efficient and strategy-proof.  
\II

For several decision problems, the only efficient and strategy-proof
payment-based mechanisms are Groves mechanisms. This is implied by a general result of
\cite{Holmstrom79:Groves}, which covers the two domains that we consider in Sections \ref{sec:multi-unit} and \ref{sec:pp-equal},
and explains our focus on Groves mechanisms. Hence from now on, we use the term ``mechanism''
to refer to a Groves mechanism. 

Focusing on the set of non-deficit Groves mechanisms, individually (respectively,
collectively) undominated mechanisms are the mechanisms from this set
that are not individually (respectively, collectively)
dominated by any other non-deficit Groves mechanism. As mentioned
earlier, no matter which domain and which set of mechanisms we
consider, collective undominance always implies individual
undominance. In Section~\ref{subsec:distinct} we show two examples of single-item auction scenarios, where collective undominance is strictly stronger than
individual undominance, for non-deficit Groves
mechanisms. That is, there exists an individually undominated
non-deficit Groves mechanism that is collectively dominated.

Recall that a special Groves mechanism, called the \bfe{VCG} or
\bfe{Clarke} mechanism~\cite{Clarke71:Multipart}, is obtained
using\footnote{Here and below, whenever $D$ is not a finite set, in
  order to ensure that the considered maximum exists, we assume that
  $f$ is continuous, and so is $v_i$ for each $i$, and also that the
  set $D$ and all $\Theta_i$ are compact subsets of some
  $\mathbb{R}^k$.}
\[ 
h_i(\theta_{-i}) :=  \max_{d \in D} \sum_{j \neq i} v_j(d,\theta_j).
\]

In this case
\[
t_i(\theta)  := \max_{d \in D} \sum_{j \neq i} v_j(d, \theta_j) - \sum_{j \neq i} v_j(f(\theta), \theta_j),
\]
which shows that the VCG mechanism is pay-only.

In what follows we introduce a slightly different notation to describe Groves mechanisms, that makes the rest of our presentation more convenient. 
First, we denote the payment function $t_i$ of the VCG mechanism by
$VCG_i$. Note now that each Groves mechanism $(f,t)$ can be defined in terms of
the VCG mechanism by setting $t_i(\theta) := VCG_i(\theta) - r_i(\theta_{-i})$,
where $r_i: \Theta_{-i} \myra \mathbb{R}$ is some function of $\theta_{-i}$.
We refer
then to $\textbf{r} := (r_1, \LL, r_n)$ as a \bfe{redistribution function}.
Hence each Groves mechanism can be identified with a redistribution function
$\textbf{r}$ and can be viewed as the VCG mechanism combined with a
redistribution. That is, under $\textbf{r}$ the
agents first participate in the VCG mechanism. Then, on top of that, 
agent $i$ also receives a redistribution amount equal to
$r_i(\theta_{-i})$. By definition, a Groves mechanism  $\textbf{r}$ is
non-deficit iff $\sum_{i = 1}^{n} VCG_i(\theta) \geq \sum_{i = 1}^{n}
r_i(\theta_{-i})$ for all $\theta \in \Theta$.


\subsection{Dominance relations}

Using the new notation above, individual and collective dominance (among
non-deficit Groves mechanisms) can be described as follows:

\begin{definition}
A non-deficit Groves mechanism $\bold{r'}$ \bfe{individually dominates} another
non-deficit Groves mechanism $\bold{r}$
if 
\begin{itemize}
\item for all $i$ and all $\theta$, 
$r'_i(\theta_{-i})\ge r_i(\theta_{-i})$,

\item for some $i$ and some $\theta$,
$r'_i(\theta_{-i}) > r_i(\theta_{-i})$.
\end{itemize} 
\end{definition}

\begin{definition}
A non-deficit Groves mechanism $\bold{r'}$ \bfe{collectively dominates} another
non-deficit Groves mechanism $\bold{r}$
if 
\begin{itemize}
\item for all $\theta$, 
$\sum_ir'_i(\theta_{-i})\ge \sum_ir_i(\theta_{-i})$,

\item for some $\theta$,
$\sum_ir'_i(\theta_{-i}) > \sum_ir_i(\theta_{-i})$.
\end{itemize} 
\end{definition}

We now consider the mechanism that results from applying the BCGC
transformation to the VCG mechanism. We refer to this as the
Bailey-Cavallo mechanism or simply the \bfe{BC mechanism}~\cite{Cavallo06:Optimal}. The VCG mechanism is characterized by
the constant redistribution function $\bold{r^{VCG}}=(0,0,\LL,0)$.
After the BCGC transformation, every agent $i$ receives an additional
amount of $\frac1n$ times the surplus guarantee
$S^{BCGC}_i(\theta_{-i})$, independent of his own type.  That is, the
BC mechanism is also a Groves mechanism, and its redistribution
function is $\bold{r^{BC}}=$\\
$(\frac1n S^{BCGC}_1, \frac1n S^{BCGC}_2,\LL, \frac1n S^{BCGC}_n)$.

Let $\theta' :=
(\theta_1,\LL,\theta_{i-1}, \theta_i', \theta_{i+1},\LL,\theta_n)$.
Then starting from the VCG mechanism, we have
\begin{small}
\[
S^{BCGC}_i(\theta_{-i}) = \inf_{\theta_i' \in \Theta_i} \sum_{k=1}^n \left[    
\max_{d \in D} \sum_{j \neq k} v_j(d, \theta'_j) - \sum_{j \neq k} v_j(f(\theta'), \theta'_j)
\right],
\]
\end{small}
that is,
\begin{small}
\begin{equation}
 \label{eq:BC}
 S^{BCGC}_i(\theta_{-i}) = \inf_{\theta_i'\in\Theta_i} \left[ 
 \sum_{k=1}^n \max_{d \in D} \sum_{j \neq k} v_j(d, \theta'_j) - (n-1) \sum_{k=1}^n  v_k(f(\theta'), \theta'_k)
\right]
\end{equation}
\end{small}
In many settings, we have that for all $\theta$ and for all $i$,
$S^{BCGC}_i(\theta_{-i}) = 0$,
and consequently the VCG and BC mechanisms coincide ({\em e.g.}, see Proposition~\ref{prop:BC-VCG}).
Whenever they do not,
by Proposition~\ref{prop:bc}$(ii)$, BC individually and collectively dominates VCG.  This is the case for the
single-item auction, as it can be seen that there $S^{BCGC}_i(\theta_{-i}) =
[{\theta_{-i}]}_2$, where $[\theta_{-i}]_2$ is the
second-highest bid among bids other than agent $i$'s own bid.

\subsection{Anonymous Groves mechanisms}
\label{subsec:anonymous}
Some of the proofs of our main results are obtained by arguing first about 
a special class
of Groves mechanisms, called anonymous Groves mechanisms.
We provide here some results about this class that we will utilize in later
sections.  We call a function $f: A^n \myra B$ \bfe{permutation
  independent} if for all permutations $\pi$ of $\{1, \LL, n\}$, $f = f
\circ \pi$.
Following \cite{Moulin86:Characterization}, we call a Groves mechanism 
$\textbf{r} = (r_1, \LL, r_n)$ \bfe{anonymous}\footnote{Our definition is 
slightly different than the one introduced in \cite{Moulin86:Characterization} in that no
conditions are put on the utility functions and the permutation
independence refers only to the redistribution function.} if
\begin{itemize}
\item all type sets $\Theta_i$ are equal,

\item all functions $r_i$ coincide and each of them is permutation independent.
\end{itemize}
Hence, an anonymous Groves mechanism is uniquely determined by a single
function $r : \Theta^{n-1} \rightarrow \mathbb R$.

In general, the VCG mechanism is not anonymous. But it is anonymous when
all the type sets are equal and all the initial utility functions $v_i$
coincide. This is the case in the two domains that we consider in later sections.

For any $\theta \in \Theta$ and any permutation $\pi$ of $\{1, \LL, n\}$ we define $\theta^{\pi} \in \Theta$ by letting
\[
\theta^{\pi}_i := \theta_{\pi^{-1}(i)}.
\]

Denote by $\Pi(k)$ the set of all permutations of the set $\{1, \LL, k\}$.
Given a Groves mechanism $\textbf{r}:= (r_1, \LL, r_n)$ for which the type set $\Theta_i$ is the same
for every agent (and equal to some set $\Theta_0$),
we construct now a function $r': \Theta^{n-1}_0 \myra \mathbb R$, following \cite{Moulin86:Characterization}, by setting 
\[
r'(x) := \sum_{j = 1}^{n} \frac{\sum_{\pi \in \Pi(n-1)} r_j(x^{\pi})}{n!},
\]
where $x^{\pi}$ is defined analogously to $\theta^{\pi}$.

Note that $r'$ is permutation independent, so $r'$ is an anonymous Groves mechanism.
The following lemma, which can be of independent interest, shows that some of the properties of $\textbf{r}$ transfer to $r'$.

\begin{lemma} 
\label{lem:anon}
Consider a Groves mechanism $\textbf{r}$ and the corresponding anonymous Groves mechanism $r'$. 
Let $VCG(\theta) := \sum_{i = 1}^{n}  VCG_i(\theta)$, and suppose that the $VCG$ function is permutation independent.
Then:
  \begin{enumerate} \smallromani
  \item  If $\textbf{r}$ is non-deficit, so is $r'$.

  \item   If an anonymous Groves mechanism $r^0$ is collectively dominated by $\textbf{r}$, 
then it is collectively dominated by $r'$.
  \end{enumerate}
\end{lemma}

\begin{Proof}
For all $\theta \in \Theta$ we have 
\[
\sum_{i = 1}^{n} r'_i(\theta_{-i}) = \frac{\sum_{i = 1}^{n} \sum_{\pi \in \Pi(n-1)} \sum_{j = 1}^{n} r_j((\theta_{-i})^{\pi})}{n!} =
\]
\[
\frac{\sum_{i=1}^n\sum_{\pi \in \Pi(n)} r_{i}(\theta^{\pi}_{-i})}{n!}
\]
where the last equality holds since in both terms we aggregate over
all applications of all $r_i$ functions to all permutations of $n-1$
elements of $\theta$.

Let $t$ and $t'$ be the payment functions of the mechanisms $\textbf{r}$ and $r'$, respectively. We have
\[
\sum_{i = 1}^{n} t_i'(\theta)
= VCG(\theta) - \sum_{i = 1}^{n} r'_i(\theta_{-i})
\]
and for all $\pi \in \Pi(n)$
\[
\sum_{i = 1}^{n} t_i(\theta^{\pi})
= VCG(\theta^{\pi}) - \sum_{i = 1}^{n} r_i(\theta^{\pi}_{-i}).
\]


Hence by the assumption about $VCG(\theta)$ it follows that

\begin{equation}
  \label{equ:ano}
\sum_{i = 1}^{n} t'_i(\theta) = 
\frac{\sum_{\pi \in \Pi(n)} \sum_{i = 1}^{n} t_i(\theta^{\pi})}{n!}
\end{equation}
$(i)$ is now an immediate consequence of (\ref{equ:ano}).

To prove $(ii)$ let $t^0$ be the payment function of $r^0$. 
$\textbf{r}$ collectively dominates $r^0$, so for all $\theta \in \Theta$ and all $\pi \in \Pi(n)$
\[
\sum_{i = 1}^{n} t_i(\theta^{\pi}) \leq \sum_{i = 1}^{n} t^0_i(\theta^{\pi})
\]
with at least one inequality strict.
Hence for all $\theta \in \Theta$
\[
\frac{\sum_{\pi \in \Pi(n)} \sum_{i = 1}^{n} t_i(\theta^{\pi})}{n!} \leq \frac{\sum_{\pi \in \Pi(n)} \sum_{i = 1}^{n} t^0_i(\theta^{\pi})}{n!}
\]
with at least one inequality strict.

But the fact that $r^0$ is anonymous and the assumption about $VCG(\theta)$ imply that
for all $\theta \in \Theta$ and all permutations $\pi$ of $\{1, \LL, n\}$
\[
\sum_{i = 1}^{n} t^0_i(\theta^{\pi}) = \sum_{i = 1}^{n} t^0_i(\theta),
\]
so by (\ref{equ:ano}) and the above inequality, we have that for all $\theta \in \Theta$
\[
\sum_{i = 1}^{n} t'_i(\theta) \leq
\sum_{i = 1}^{n} t^0_i(\theta),
\]
with at least one inequality strict.
\end{Proof}

The assumption in
Lemma~\ref{lem:anon} of permutation independence of $VCG(\theta)$ is satisfied in
both of the domains that we consider in Sections \ref{sec:multi-unit} and \ref{sec:pp-equal}.
So item $(ii)$ states that if a Groves mechanism considered in the sequel
is not collectively undominated, then it is collectively dominated by an anonymous Groves
mechanism. 

We now prove that for a large class of Groves mechanisms that includes the ones
we study in the sequel the introduced relations of dominance differ.

\begin{theorem} \label{thm:diff}
Suppose $n \geq 3$.
  Assume that the sets of types $\Theta_i$ are all equal to the set $\Theta_0$ which
contains at least $n-1$ elements. Then two non-deficit anonymous
Groves mechanisms $r$ and $r'$ exist such that $r$ collectively dominates $r'$
but $r$ does not individually dominate $r'$.
\end{theorem}

\begin{Proof}
 Fix a non-deficit anonymous
Groves mechanism determined by a permutation independent function $r: \Theta^{n-1}_0 \myra \mathbb R$.

Let $a_1, \LL, a_{n-1}$ be arbitrary different elements of $\Theta_0$.
Define a permutation independent function $q: \Theta^{n-1}_0 \myra \mathbb R$ by putting
\[
        q(x) :=
        \left\{
        \begin{array}{l@{\extracolsep{3mm}}l}
        -1    & \mbox{if $x$ is a permutation of $(a_1, \LL, a_{n-1})$} \\
        \phantom{-}2       & \mathrm{otherwise}
        \end{array}
        \right.
\]

Then for each $\theta \in \Theta^{n}_0$ at most two of its subsequences $\theta_{-i}$
may form a permutation of $(a_1, \LL, a_{n-1})$. But $n \geq 3$, so 
for all $\theta \in \Theta$,
$
\sum_{i = 1}^{n} q(\theta_{-i}) \geq 0
$.
This implies that the anonymous
Groves mechanism determined by the function 
$r' := r - q$ is non-deficit. 

Trivially, the sum of payments under $r$ is less than or equal to the sum of payments
under $r'$, since $r$ redistributes more money than $r'$. 
Moreover for some $\theta \in \Theta$, for instance $\theta = (a_1, \LL, a_1)$, we have
$
\sum_{i = 1}^{n} q(\theta_{-i}) > 0
$.
Finally, by definition, $q(a_1, \LL, a_{n-1}) = -1$.

These imply that $r'$ is collectively dominated by $r$ but
is not individually dominated by $r$.
\end{Proof}

\section{Individually undominated mechanisms: characterization and algorithmic techniques for general domains}
\label{sec:individual}

In this section, we focus on individually undominated non-deficit Groves
mechanisms.

\subsection{Non-deficit Groves mechanisms}

We start with a characterization of non-deficit Groves mechanisms.
Recall first that for a type profile $\theta$, we denote by
$VCG(\theta)$ the total VCG payment, $\sum_{i=1}^n VCG_i(\theta)$.
\begin{proposition} 
    \label{prop:feasible} 
    A Groves mechanism $\bold{r}$ is {\em
    non-deficit} if and only if for all $i$ and all $\theta$,
    \begin{equation}
    \label{eq:feasible}
        r_i(\theta_{-i})\le \inf\limits_{\theta'_i \in \Theta_i}
        \{VCG(\theta'_i,\theta_{-i})-\sum\limits_{j\ne i}r_j(\theta'_{-j})\} 
    \end{equation}
Here, $\theta'_{-j}$ are the reported types of the agents other
than $j$ when $\theta_i$ is replaced by $\theta'_i$.
\end{proposition}

\begin{Proof} We first prove the ``if'' direction.  For any $i$ and $\theta$,
Equation~\ref{eq:feasible} implies that $r_i(\theta_{-i})\le
VCG(\theta'_i,\theta_{-i})-\sum\limits_{j\ne i}r_j(\theta'_{-j})$ for any
$\theta'_i\in \Theta_i$. If we let $\theta'_i=\theta_i$, we obtain
$\sum_jr_j(\theta_{-j}) \le VCG(\theta_i, \theta_{-i})=\sum_iVCG_i(\theta)$.
Thus, the non-deficit property holds.

We now prove the ``only if'' direction.  
To ensure the non-deficit property, for any $i$, any $\theta_{-i}$, and any $\theta'_i$, we must have
$r_i(\theta_{-i}) + \sum\limits_{j\ne i}r_j(\theta'_{-j}) \leq
VCG(\theta'_i,\theta_{-i})$, or equivalently $r_i(\theta_{-i})\le
VCG(\theta'_i,\theta_{-i})-\sum\limits_{j\ne i}r_j(\theta'_{-j})$.  Since
$\theta'_i$ is arbitrary, Equation~\ref{eq:feasible} follows.\end{Proof}

By replacing the ``$\le$'' in Equation~\ref{eq:feasible} by ``$=$'', we get a
characterization of individually undominated non-deficit Groves mechanisms.

\begin{theorem} 
\label{th:characterization}
A Groves mechanism $\bold{r}$ is non-deficit and individually undominated if and
only if for all $i$ and all $\theta$,
\begin{equation}
    \label{eq:undominated} 
    r_i(\theta_{-i}) = \inf\limits_{\theta'_i \in \Theta_i}
        \{VCG(\theta'_i,\theta_{-i})-\sum\limits_{j\ne i}r_j(\theta'_{-j})\} 
\end{equation}
Here, $\theta'_{-j}$ are the reported types of the agents other
than $j$ when $\theta_i$ is replaced by $\theta'_i$.
\end{theorem}
\begin{Proof}
We prove the ``if'' direction first.  
Any Groves mechanism
$\bold{r}$ that satisfies 
Equation~\ref{eq:undominated} is non-deficit by Proposition~\ref{prop:feasible}. Now
suppose that $\bold{r}$ is individually dominated, that is, there exists another non-deficit
Groves mechanism $\bold{r'}$ such that for all $i$ and
$\theta_{-i}$, we have $r'_i(\theta_{-i})\ge r_i(\theta_{-i})$, and for
some $i$ and $\theta_{-i}$, we have $r'_i(\theta_{-i}) >
r_i(\theta_{-i})$. For the $i$ and $\theta_{-i}$ that make this inequality
strict, we have
\[r'_i(\theta_{-i})> r_i(\theta_{-i}) =
\inf\limits_{\theta'_i\in
\Theta_i}\{VCG(\theta'_i,\theta_{-i})-\sum\limits_{j\ne
i}r_j(\theta'_{-j})\}\]
\[\ge \inf\limits_{\theta'_i\in
\Theta_i}\{VCG(\theta'_i,\theta_{-i})-\sum\limits_{j\ne
i}r'_j(\theta'_{-j})\},\]
which contradicts with the fact that $\bold{r}'$ must satisfy Equation~\ref{eq:feasible}.  It follows that $\bold{r}$ is individually undominated.

Now we prove the ``only if'' direction.  
Suppose Equation~\ref{eq:undominated} is
not satisfied.  Then, there exists some $i$ and $\theta_{-i}$ such that
$r_i(\theta_{-i}) < \inf\limits_{\theta'_i\in
\Theta_i}\{VCG(\theta'_i,\theta_{-i})-\sum\limits_{j\ne
i}r_j(\theta'_{-j})\}$.  Let $a = \inf\limits_{\theta'_i\in
\Theta_i}\{VCG(\theta'_i,\theta_{-i})-\sum\limits_{j\ne
i}r_j(\theta'_{-j})\} - r_i(\theta_{-i})$ (so that $a>0$), and let
$\bold{r'}$ be the same as $\bold{r}$, except that for the aforementioned
$i$ and $\theta_{-i}$, $r'_i(\theta_{-i})= r_i(\theta_{-i})+ a$.  To show
that this does not break the non-deficit constraint, consider any type
vector $(\theta_i, \theta_{-i})$ where $i$ and $\theta_{-i}$ are the same
as before (that is, any type profile that is affected).  Then,
\[r'_i(\theta_{-i})=a+r_i(\theta_{-i}) =\inf\limits_{\theta'_i\in
\Theta_i}\{VCG(\theta'_i,\theta_{-i})-\sum\limits_{j\ne
i}r_j(\theta'_{-j})\}\] 
\[= \inf\limits_{\theta'_i\in
\Theta_i}\{VCG(\theta'_i,\theta_{-i})-\sum\limits_{j\ne
i}r'_j(\theta'_{-j})\}.\]  
Thus, by Proposition~\ref{prop:feasible}, $\bold{r'}$ is
non-deficit.  This contradicts that $\bold{r}$ is individually undominated.  Hence,
Equation~\ref{eq:undominated} must hold.
\end{Proof}

We now give an
example of an individually undominated mechanism.

\begin{example} Consider a single-item auction with $n \geq 3$ agents. Agent
$i$ bids $\theta_i \in [0,\infty)$.  Let $[\theta]_j$ be the $j$th highest type
from the type profile $\theta$.  Let us consider the anonymous Groves mechanism
characterized by $r(\theta_{-i})=\frac{1}{n}[\theta_{-i}]_2$.  That is, under
this mechanism, besides paying the VCG payment, every agent receives $\frac1n$
times the second highest {\em other} bid.  In fact, this mechanism is the BC
mechanism for single-item auctions. To show that $r$ is individually undominated, it
suffices to show Equation~\ref{eq:undominated} is satisfied.  We first observe
that for every agent, the second highest {\em other} bid is no more than the
second highest bid, which equals the total VCG payment.  That is, $r$ is
non-deficit. Hence, Equation~\ref{eq:feasible} holds for all agents and all
type profiles.  Moreover, for every type profile $\theta$, by setting
$\theta'_i = [\theta_{-i}]_2$, we can verify that Equation~\ref{eq:undominated}
holds.  It follows that the BC mechanism is individually undominated for
single-item auctions.  \end{example}

In what follows, we first show two examples of single-item auction
scenarios, where collective
undominance is strictly stronger than individual undominance for non-deficit Groves mechanisms. 
We then propose
two techniques for generating individually undominated mechanisms starting from
known individually dominated mechanisms (if the initial mechanism is already
individually undominated, then the techniques will return the same mechanism).
One technique immediately produces an individually undominated mechanism.
However, it does not preserve anonymity.  The second technique preserves
anonymity, and after repeated applications the result converges to an
individually undominated mechanism.  We emphasize that we can start with {\em
any} non-deficit Groves mechanism, including the BC mechanism, the Worst-Case
Optimal mechanism~\cite{Guo07:Worst}, the Optimal-in-Expectation Linear
mechanisms~\cite{Guo10:Optimal}, and the VCG mechanism.

\subsection{Collective undominance is strictly stronger than individual undominance}
\label{subsec:distinct}

We use two examples to show that collective undominance is, in general, strictly
stronger than individual undominance.

\begin{example} Consider a single-item auction with $4$ agents.  We assume that
for each agent, the set of allowed types is the same, namely, integers from $0$
to $3$.  Here, the VCG mechanism is just the second-price auction. 

Let us consider the following two anonymous non-deficit Groves mechanisms, which are
computer-generated for differentiating collective undominance and individual undominance.
\\
\\
{\bf Mechanism 1:} 
$r(\theta_{-i})=r([\theta_{-i}]_1,[\theta_{-i}]_2,[\theta_{-i}]_3)$, and
the function $r$ is given in Table~\ref{tb:example}. ($[\theta_{-i}]_j$ is the $j$th highest type among types other than $i$'s own type.)
\\
\\
{\bf Mechanism 2:}
$r'(\theta_{-i})=r'([\theta_{-i}]_1,[\theta_{-i}]_2,[\theta_{-i}]_3)$, and
the function $r'$ is given in Table~\ref{tb:example}.
\\

\begin{table}
        \begin{center}
\begin{tabular}{|c|c|c|c|}
\hline
$\mathbf{r(0,0,0)}$ & $0$ & $\mathbf{r'(0,0,0)}$ & $0$\\
\hline
$\mathbf{r(1,0,0)}$ & $0$ & $\mathbf{r'(1,0,0)}$ & $0$\\
\hline
$\mathbf{r(1,1,0)}$ & $1/4$ & $\mathbf{r'(1,1,0)}$ & $1/4$\\
\hline
$\mathbf{r(1,1,1)}$ & $1/4$ & $\mathbf{r'(1,1,1)}$ & $1/4$\\
\hline
$\mathbf{r(2,0,0)}$ & $0$ & $\mathbf{r'(2,0,0)}$ & $0$\\
\hline
$\mathbf{r(2,1,0)}$ & $1/12$ & $\mathbf{r'(2,1,0)}$ & $7/24$\\
\hline
$\mathbf{r(2,1,1)}$ & $0$ & $\mathbf{r'(2,1,1)}$ & $1/6$\\
\hline
$\mathbf{r(2,2,0)}$ & $1/2$ & $\mathbf{r'(2,2,0)}$ & $1/2$\\
\hline
$\mathbf{r(2,2,1)}$ & $0$ & $\mathbf{r'(2,2,1)}$ & $1/4$\\
\hline
$\mathbf{r(2,2,2)}$ & $1/2$ & $\mathbf{r'(2,2,2)}$ & $1/2$\\
\hline
$\mathbf{r(3,0,0)}$ & $0$ & $\mathbf{r'(3,0,0)}$ & $0$\\
\hline
$\mathbf{r(3,1,0)}$ & $1/4$ & $\mathbf{r'(3,1,0)}$ & $1/4$\\
\hline
$\mathbf{r(3,1,1)}$ & $0$ & $\mathbf{r'(3,1,1)}$ & $1/4$\\
\hline
$\mathbf{r(3,2,0)}$ & $2/3$ & $\mathbf{r'(3,2,0)}$ & $2/3$\\
\hline
$\mathbf{r(3,2,1)}$ & $1$ & $\mathbf{r'(3,2,1)}$ & $19/24$\\
\hline
$\mathbf{r(3,2,2)}$ & $0$ & $\mathbf{r'(3,2,2)}$ & $1/6$\\
\hline
$\mathbf{r(3,3,0)}$ & $2/3$ & $\mathbf{r'(3,3,0)}$ & $5/6$\\
\hline
$\mathbf{r(3,3,1)}$ & $0$ & $\mathbf{r'(3,3,1)}$ & $7/12$\\
\hline
$\mathbf{r(3,3,2)}$ & $1$ & $\mathbf{r'(3,3,2)}$ & $5/6$\\
\hline
$\mathbf{r(3,3,3)}$ & $0$ & $\mathbf{r'(3,3,3)}$ & $1/2$\\
\hline
\end{tabular}
\end{center}
\caption{Computer-generated example mechanisms for differentiating collective undominance and individual undominance.
\label{tb:example}}
\end{table}

With the above characterization, we have that mechanism 2 collectively
dominates mechanism 1: for example, for the type profile $(3,2,2,2)$,
$\sum_ir(\theta_{-i})=1/2 < 1=\sum_ir'(\theta_{-i})$.  On the other hand,
mechanism 2 does not individually dominate mechanism 1: for example, $r(3,3,2)
= 1 > 5/6 = r'(3,3,2)$.  In fact, based on the characterization of individually
undominated non-deficit Groves mechanisms
(Theorem~\ref{th:characterization}), we are able to show that mechanism 1 is
individually undominated. \end{example}

\begin{example}
Consider a single-item auction with $5$ agents.  We assume that
for each agent, the set of allowed types is $[0,\infty)$.  Here, the VCG mechanism is just the second-price auction. 

Let us consider the following two anonymous non-deficit Groves mechanisms:
\\
\\
{\bf Mechanism 1:} 

$r(\theta_{-i})=0$ if all four types in $\theta_{-i}$ are identical.

$r(\theta_{-i})=[\theta_{-i}]_1/4$ if the highest three types in $\theta_{-i}$ are identical, and they are strictly higher than the lowest type in $\theta_{-i}$.

$r(\theta_{-i})=[\theta_{-i}]_1/6$ if the highest two types in $\theta_{-i}$ are identical, and they are strictly higher than the third highest type in $\theta_{-i}$.

$r(\theta_{-i})=3[\theta_{-i}]_2/16$ if the highest type in $\theta_{-i}$ is strictly higher than the second highest type in $\theta_{-i}$, and the second highest type in $\theta_{-i}$ is identical to the third highest type in $\theta_{-i}$.

$r(\theta_{-i})=[\theta_{-i}]_2/5$ if the highest three types in $\theta_{-i}$ are all different.
\\
\\
{\bf Mechanism 2 (BC):} 

$r'(\theta_{-i})=[\theta_{-i}]_2/5$. 
\\

With the above characterization, we have that mechanism 2 collectively
dominates mechanism 1:
for example, for the type profile $(3,2,2,2,2)$, $\sum_ir(\theta_{-i})=4r(3,2,2,2)+r(2,2,2,2)=3/2+0=3/2 <
\sum_ir'(\theta_{-i})=4r'(3,2,2,2)+r'(2,2,2,2)=8/5+2/5=2$.  On the other hand, mechanism 2 does not individually
dominate mechanism 1: for example, $r(4,4,4,1) = 1 > 4/5 = r'(4,4,4,1)$.  
In fact, based on the characterization of individually
undominated non-deficit Groves mechanisms
(Theorem~\ref{th:characterization}), we are able to show that mechanism 1 is
individually undominated. 
\end{example}

\subsection{A priority-based technique}
\label{subsec:priority}

Given a non-deficit Groves mechanism $\bold{r}$ and a priority order over
agents $\pi$, we can improve $\bold{r}$ into an individually undominated
mechanism as follows:

\vspace{0.1in} 1) Let $\pi:\{1,\ldots,n\}\rightarrow\{1,\ldots,n\}$ be a
permutation representing the priority order. That is, $\pi(i)$ is agent $i$'s
priority value (the lower the value, the higher the priority).  $\pi^{-1}(k)$
is then the agent with the $k$th highest priority.
The high-level idea of the priority-based technique is that we go over the
agents one by one. For the first agent (the agent with the highest priority),
we maximize his redistribution function subject to the constraint of
Proposition~\ref{prop:feasible}. For later agents, we do the same, but take
into consideration that earlier agents' redistribution functions have been
updated.  A priority order can be arbitrary. Generally, agents with high
priorities benefit more from this technique, since for earlier agents, there
is more room for improvement.

\vspace{0.1in} 2) Let $i=\pi^{-1}(1)$, and update $r_i$ to
\[r^{\pi}_i(\theta_{-i})=\inf\limits_{\theta'_i\in \Theta_i} \{
VCG(\theta'_i,\theta_{-i})-\sum\limits_{\pi(j)>1}r_j(\theta'_{-j})\}.\]  That
is, the update ensures that at this point $\bold{r^{\pi}}$ satisfies
Equation~\ref{eq:undominated} for $i=\pi^{-1}(1)$.  

It should be noted that during the above update, only the payment of agent $i=\pi^{-1}(1)$ 
is changed, and it is changed by
\[r^{\pi}_i(\theta_{-i})-r_i(\theta_{-i})=\inf\limits_{\theta'_i\in \Theta_i} \{
VCG(\theta'_i,\theta_{-i})-\sum\limits_{\pi(j)>1}r_j(\theta'_{-j})\}
-r_i(\theta_{-i})\]
\[=\inf\limits_{\theta'_i\in \Theta_i} \{
VCG(\theta'_i,\theta_{-i})-\sum\limits_{j}r_j(\theta'_{-j})\}
=S_i^{BCGC}(\theta_{-i}).\]
That is, essentially, the above update amounts to applying the $BCGC(i)$ transformation on $\bold{r}$, where $i=\pi^{-1}(1)$.

\vspace{0.1in} 3) We will now consider the remaining agents in turn, according
to the order $\pi$.  In the $k$th step, we update $r_i$ ($i=\pi^{-1}(k)$) to
\[r^{\pi}_i(\theta_{-i})=\inf\limits_{\theta'_i\in \Theta_i} \{
VCG(\theta'_i,\theta_{-i})-\sum\limits_{\pi(j)>k}r_j(\theta'_{-j})-\sum\limits_{\pi(j)<k}r^{\pi}_j(\theta'_{-j})\}.\]
That is, the update ensures that at this point $\bold{r^{\pi}}$ satisfies
Equation~\ref{eq:undominated} when $i=\pi^{-1}(k)$.  To avoid breaking the
non-deficit property, when we make the update, we take the previous $k-1$
updates into account.
For this update, what we are doing is essentially
applying the $BCGC(\pi^{-1}(k))$ transformation on the resulting mechanism from
the previous update.

\vspace{0.1in} Overall, for every agent $i$,
\[r^{\pi}_i(\theta_{-i})=\inf\limits_{\theta'_i\in \Theta_i} \{
VCG(\theta'_i,\theta_{-i})-\sum\limits_{\pi(j)>\pi(i)}r_j(\theta'_{-j})-\sum\limits_{\pi(j)<\pi(i)}r^{\pi}_j(\theta'_{-j})\}.\]

The new mechanism $\bold{r^{\pi}}$ satisfies the following properties:

\vspace{0.05in} \begin{proposition} \label{prop:permuir}        For all $i$ and
$\theta_{-i}$, $r_i^{\pi}(\theta_{-i})\geq r_i(\theta_{-i})$.
\end{proposition}

\begin{Proof} First consider $i=\pi^{-1}(1)$, the agent with the highest
priority. For any $\theta_{-i}$, we have
$r^{\pi}_i(\theta_{-i})=\inf\limits_{\theta'_i\in \Theta_i} \{
VCG(\theta'_i,\theta_{-i})-\sum\limits_{j\ne i}r_j(\theta'_{-j})\}$.  Since 
$\bold{r}$ is non-deficit, by
Equation~\ref{eq:feasible}, we have $r_i(\theta_{-i})\leq
\inf\limits_{\theta'_i\in \Theta_i} \{
VCG(\theta'_i,\theta_{-i})-\sum\limits_{j\ne i}r_j(\theta'_{-j})\}$.  Hence
$r_i^{\pi}(\theta_{-i})\geq r_i(\theta_{-i})$.

For any $i\ne \pi^{-1}(1)$, $r^{\pi}_i(\theta_{-i})$ equals
\[r_i(\theta_{-i})+\inf\limits_{\theta'_i\in \Theta_i}
\{VCG(\theta'_i,\theta_{-i})
-r_i(\theta_{-i})-\sum\limits_{\pi(j)>\pi(i)}r_j(\theta'_{-j})-\sum\limits_{\pi(j)<\pi(i)}r^{\pi}_j(\theta'_{-j})\}.\]

We must show that 
\begin{equation}\inf\limits_{\theta'_i\in \Theta_i} \{
VCG(\theta'_i,\theta_{-i})-r_i(\theta_{-i})-\sum\limits_{\pi(j)>\pi(i)}r_j(\theta'_{-j})-
\sum\limits_{\pi(j)<\pi(i)}r^{\pi}_j(\theta'_{-j})\}\geq 0.\label{eq:tech1}\end{equation}

Consider $p=\pi^{-1}(\pi(i)-1)$ (the agent immediately before $i$ in terms of
priority). For any $\theta_i, \theta_{-i}$, we have
\[VCG(\theta_i,\theta_{-i})-r_i(\theta_{-i})-\sum\limits_{\pi(j)>\pi(i)}r_j(\theta_{-j})-\sum\limits_{\pi(j)<\pi(i)}r^{\pi}_j(\theta_{-j})\]
\[=VCG(\theta_i,\theta_{-i})-\sum\limits_{\pi(j)>\pi(p)}r_j(\theta_{-j})-\sum\limits_{\pi(j)<\pi(p)}r^{\pi}_j(\theta_{-j})-r^{\pi}_p(\theta_{-p})\]
\[\geq \inf\limits_{\theta'_p\in \Theta_p} \{VCG(\theta'_p,
\theta_{-p})-\sum\limits_{\pi(j)>\pi(p)}r_j(\theta'_{-j})-\sum\limits_{\pi(j)<\pi(p)}r^{\pi}_j(\theta'_{-j})\}-r^{\pi}_p(\theta_{-p})=0.\] 
In the above inequality, $\theta'_{-j}$ is the set of types reported by
the agents other than $j$, when $\theta_p$ is replaced by $\theta'_p$.
Because $\theta_i$ is arbitrary, Equation~\ref{eq:tech1} follows.
Therefore, $r_i^{\pi}(\theta_{-i})\geq r_i(\theta_{-i})$ for all $i$
and $\theta_{-i}$.  \end{Proof}

\vspace{0.05in} \begin{proposition}\label{prop:piundominated} $\bold{r^\pi}$ is
individually undominated.  \end{proposition}

\begin{Proof} 
Let $i = \pi^{-1}(n)$.  For all $\theta$, 
\[VCG(\theta)-\sum\limits_{j=1,\dots,n}r^{\pi}_j(\theta_{-j})
\geq \inf\limits_{\theta'_i\in
\Theta_i}\{VCG(\theta'_i,\theta_{-i})-\sum\limits_{j\ne
i}r^{\pi}_j(\theta'_{-j})\}-r^{\pi}_i(\theta_{-i})=0.\]
Hence $\bold{r^\pi}$ never incurs a deficit. So, $\bold{r^\pi}$ is non-deficit.

Using Proposition~\ref{prop:permuir}, we have for all $i$ and all $\theta$,
\[r^{\pi}_i(\theta_{-i})=\inf\limits_{\theta'_i\in \Theta_i} \{
VCG(\theta'_i,\theta_{-i})-\sum\limits_{\pi(j)>\pi(i)}r_j(\theta'_{-j})-\sum\limits_{\pi(j)<\pi(i)}r^{\pi}_j(\theta'_{-j})\}\]\[\ge
\inf\limits_{\theta'_i\in \Theta_i}
\{VCG(\theta'_i,\theta_{-i})-\sum\limits_{j\ne i}r^{\pi}_j(\theta'_{-j})\}.\]
Because $\bold{r^\pi}$ is non-deficit, the opposite inequality must also be
satisfied (Equation~\ref{eq:feasible})---hence we must have equality, that is,
Equation~\ref{eq:undominated} must hold.
It follows that $\bold{r^\pi}$ is individually
undominated.  \end{Proof}

It should be noted that for the above technique, during the updates, we need
to keep track of the value of $r^{\pi}_i(\theta_{-i})$ for all $i$ and
$\theta_{-i}$. That is, due to space complexity, the above technique is more
suitable for cases with few agents and few possible types.  To reduce space
complexity, when we update, we could also recompute earlier updates in a
recursive fashion. By doing so, the later updates are
much more difficult to compute compared to the earlier updates. Fortunately, the earlier updates tend to be more important, because there is generally
more room for improvement during the earlier updates. Therefore, a reasonable
approximation would be to update only for a few high-priority agents and ignore
the remaining agents with low priorities.

\subsection{An iterative technique that preserves anonymity}
\label{subsec:iterative}

The previous technique will, in general, not produce an anonymous mechanism,
even if the input mechanism is anonymous.  This is because agents higher in the
priority order tend to benefit more from the technique.  Here, we will
introduce another technique that preserves anonymity.

Given an anonymous mechanism $r$, let $r^0=r$.
For all $i$ and all
$\theta$, let
\[r^{k+1}(\theta_{-i})=
\frac{n-1}{n}r^{k}(\theta_{-i})+\frac{1}{n}\inf\limits_{\theta'_i\in
\Theta_i}\{VCG(\theta'_i,\theta_{-i})-\sum\limits_{j\ne i}r^{k}(\theta'_{-j})\}.\]

It is easily seen by induction that all the $r^k$ mechanisms are anonymous. If $r^k$ is anonymous, then for any
$\pi\in \Pi(n-1)$, $r^{k}((\theta_{-i})^{\pi})=r^{k}(\theta_{-i})$ for all
$\theta$ and all $i$. We also have that
$VCG(\theta'_i,(\theta_{-i})^{\pi})=VCG(\theta'_i,\theta_{-i})$ for all
$\theta$, all $\theta'_i$, and all $i$.  Finally, let
$((\theta_{-i})^{\pi},\theta'_i)$ be the type profile where the types in
$\theta_{-i}$ are permuted according to $\pi$, and $\theta_i$ is replaced by
$\theta'_i$. We have
$\sum_{j\neq i}r^{k}(((\theta_{-i})^{\pi},\theta'_i)_{-j})=\sum_{j\neq i}r^{k}(\theta'_{-j})$ for all
$\theta$, all $i$, and all $\theta'_i$.  The above implies that
$r^{k+1}$ is also permutation independent, thus anonymous.

It should be noted that from $r^k$ to $r^{k+1}$, agent $i$'s payment
is changed by
\begin{eqnarray*}
\phantom{=}& r^{k+1}(\theta_{-i})-r^{k}(\theta_{-i}) \\
          =& \frac{n-1}{n}r^{k}(\theta_{-i})+\frac{1}{n}\inf\limits_{\theta'_i\in
\Theta_i}\{VCG(\theta'_i,\theta_{-i})-\sum\limits_{j\ne i}r^{k}(\theta'_{-j})\}-r^k(\theta_{-i}) \\
          =& \frac{1}{n}\inf\limits_{\theta'_i\in
\Theta_i}\{VCG(\theta'_i,\theta_{-i})-\sum\limits_{j}r^{k}(\theta'_{-j})\} \\
          =& \frac{1}{n}S_i^{BCGC}(\theta_{-i}).
\end{eqnarray*}
That is, essentially, $r^{k+1}$ is the resulting mechanism by applying the $BCGC$ transform on $r^k$.


The next propositions immediately follow from Proposition~\ref{prop:bc}:

\begin{proposition} If $r^0$ is non-deficit, then $r^{k}$ is non-deficit for all $k$.
\label{prop:iterative_feasible}
\end{proposition}


\begin{proposition}
For all $i$ and $\theta_{-i}$,
$r^k(\theta_{-i})$ is nondecreasing in $k$.
\label{prop:increasing_in_k}
\end{proposition}


\begin{proposition} If $r^{k+1}=r^k$, then $r^k$ is individually undominated.
\end{proposition}

\begin{proposition} \label{prop:last} If $r^{k}$ is not individually undominated, then $r^{k+1}$ individually dominates $r^k$.
\end{proposition}


Finally, the following proposition establishes convergence.

\begin{proposition} \label{prop:converge}
As $k \rightarrow \infty$, $r^k$ converges (pointwise) to an individually undominated mechanism.
\end{proposition}

\begin{Proof}
By Proposition~\ref{prop:increasing_in_k}, the $r^{k}(\theta_{-i})$ are
nondecreasing in $k$, and since every $r^k$ is non-deficit by
Proposition~\ref{prop:iterative_feasible}, they must be bounded; hence they must
converge (pointwise).  For any $i$ and $\theta_{-i}$, let
\[d_k=\inf\limits_{\theta'_i\in \Theta_i}\{
VCG(\theta'_i,\theta_{-i})-\sum\limits_{j\ne
i}r^{k}(\theta'_{-j})\}-r^{k}(\theta_{-i}).\]  Using
Proposition~\ref{prop:increasing_in_k}, we derive the following inequality:
\[d_{k+1}=\inf\limits_{\theta'_i\in \Theta_i}\{
VCG(\theta'_i,\theta_{-i})-\sum\limits_{j\ne
i}r^{k+1}(\theta'_{-j})\}-r^{k+1}(\theta_{-i})\]
\[\le
\inf\limits_{\theta'_i\in \Theta_i}\{ VCG(\theta'_i,\theta_{-i})-\sum\limits_{j\ne
i}r^{k}(\theta'_{-j})\}
-r^{k+1}(\theta_{-i})\]
\[=\inf\limits_{\theta'_i\in
\Theta_i}\{VCG(\theta'_i,\theta_{-i})-\sum\limits_{j\ne i}r^{k}(\theta'_{-j})\}\] 
\[= -
\frac{n-1}{n}r^{k}(\theta_{-i})
-\frac{1}{n}\inf\limits_{\theta'_i\in
\Theta_i}\{VCG(\theta'_i,\theta_{-i})-\sum\limits_{j\ne i}r^{k}(\theta'_{-j})\}
\]
\[=\frac{n-1}{n}\inf\limits_{\theta'_i\in
\Theta_i}\{
VCG(\theta'_i,\theta_{-i})-\sum\limits_{j\ne i}r^{k}(\theta'_{-j})\} -
\frac{n-1}{n}r^{k}(\theta_{-i}) =\frac{n-1}{n}d_k.\]  
As
$k\rightarrow\infty$, $d_k =\inf\limits_{\theta'_i\in \Theta_i}\{
VCG(\theta'_i,\theta_{-i})-\sum\limits_{j\ne
i}r^{k}(\theta'_{-j})\}-r^{k}(\theta_{-i}) 
\rightarrow 0$. So in the limit,
Equation~\ref{eq:undominated} is satisfied. Thus, $r^k$ converges
(pointwise, linearly) to an individually undominated mechanism.
\end{Proof}

Similar to the priority-based technique, in the above iterative process, when
computing for $r^k$, we need the value of $r^{k-1}(\theta_{-i})$, for
all $\theta_{-i}$. That is, due to space complexity, the above technique is
more suitable for cases with few agents and few possible types.  To reduce
the space complexity we could also recompute $r^{k-1}$ in a recursive fashion. By
doing so, $r^k$ becomes much more difficult to compute for large values of $k$.
Fortunately, the earlier iterative steps are more crucial,
because there is generally more room for improvement during the earlier steps.
Therefore a reasonable approximation would be to only compute a few iterative
steps.

\section{Multi-unit auctions with unit demand} \label{sec:multi-unit}

In this section, we consider auctions where there are multiple identical units
of a single good and all agents have unit demand, {\em i.e.}, each agent wants
only one unit (if there is a single unit of the good, we simply have the
standard single-item auction). We focus on the notion of collectively
undominated mechanisms and how it relates to that of individually undominated
mechanisms. In particular, we first obtain an analytical characterization of
all collectively undominated Groves mechanisms that are non-deficit, anonymous,
and have linear payment functions, by proving that the Optimal-in-Expectation
Linear redistribution mechanisms (OEL)~\cite{Guo10:Optimal}, which include the
BC mechanism, are the only collectively undominated Groves mechanisms that are
anonymous and linear. We then show that individual undominance and collective
undominance are equivalent if we restrict our consideration to Groves
mechanisms that are anonymous and linear in the setting of multi-unit auctions
with unit demand. Note that even for single-item auctions, the examples given
in Section~\ref{subsec:distinct} show that this equivalence does not hold if we do not restrict
ourselves to linear and anonymous mechanisms.

If one mechanism collectively dominates another mechanism, then under the first
mechanism, the agents' expected total utility, if there was a prior
distribution over the agents' valuations, must be no less than that under the second mechanism, and strictly higher under minimal conditions on the prior distribution.
Therefore, a good direction in which to look
for collectively undominated mechanisms is to start with those mechanisms that
are optimal-in-expectation.

The {\em Optimal-in-Expectation Linear (OEL) redistribution mechanisms}~\cite{Guo10:Optimal}, described below, are special cases of
non-deficit Groves mechanisms that are anonymous and linear. The OEL mechanisms
are defined only for multi-unit auctions with unit demand. In a unit demand
multi-unit auction, there are $m$ indistinguishable units for sale, and each
agent is interested in only one unit. For agent $i$, his type $\theta_i$ is his
valuation for winning one unit. We assume all bids (announced types) are
bounded below by $L$ and above by $U$, {\em i.e.}, $\Theta_i = [L,U]$ (note
that $L$ can be $0$).

A {\em linear} and anonymous Groves mechanism is characterized by a function $r$ of the following form:
$r(\theta_{-i})=c_0+\sum\limits_{j=1}^{n-1}c_j[\theta_{-i}]_j$ (where
$[\theta_{-i}]_j$ is the $j$th highest bid among $\theta_{-i}$).  For OEL mechanisms, the
$c_j$'s are chosen according to one of the following options (indexed by integer parameter $k$,
where $k$ ranges from $0$ to $n$, and $k-m$ is odd):\\

%
%

$\mathbf{k=0}$: 

\[c_i=(-1)^{m-i}{n-i-1 \choose n-m-1}/{m-1 \choose i-1} \;\text{for}\; i=1,\ldots,m,\] 

\[c_0=Um/n-U\sum_{i=1}^m(-1)^{m-i}{n-i-1 \choose n-m-1}/{m-1\choose i-1},\] 

\[c_i=0 \;\text{for other values of}\; i.\]

$\mathbf{k=1,2,\ldots,m}$:

\[c_i=(-1)^{m-i}{n-i-1 \choose n-m-1}/{m-1 \choose i-1} \;\text{for}\; i=k+1,\ldots,m,\]

\[c_k=m/n-\sum_{i=k+1}^m(-1)^{m-i}{n-i-1 \choose n-m-1}/{m-1\choose i-1},\] 
\[c_i=0 \;\text{for other values of}\; i.\]

$\mathbf{k=m+1,m+2,\ldots,n-1}$: 

\[c_i=(-1)^{m-i-1}{i-1\choose
            m-1}/{n-m-1\choose n-i-1} \;\text{for}\; i=m+1,\ldots,k-1,\]

\[c_k=m/n-\sum_{i=m+1}^{k-1}(-1)^{m-i-1}{i-1\choose
            m-1}/{n-m-1\choose n-i-1},\] 
\[c_i=0 \;\text{for other values of}\; i.\]

$\mathbf{k=n}$: 

\[c_i=(-1)^{m-i-1}{i-1\choose
            m-1}/{n-m-1\choose n-i-1} \;\text{for}\; i=m+1,\ldots,n-1,\]

\[c_0=Lm/n-L\sum_{i=m+1}^{n-1}(-1)^{m-i-1}{i-1\choose
            m-1}/{n-m-1\choose n-i-1},\] 

\[c_i=0 \;\text{for other values of}\; i.\]

For example, when $k=m+1$, we have $c_{m+1}=m/n$ and $c_i=0$ for all other $i$.
For this specific OEL mechanism,
$r(\theta_{-i})=\frac{m}{n}[\theta_{-i}]_{m+1}$.  That is, besides
participating in the VCG mechanism, every agent also receives an amount that is
equal to $m/n$ times the $(m+1)$th highest bid from the other agents.
Actually, this is exactly the BC mechanism for multi-unit auctions with
unit demand.

Besides being non-deficit, one property of the OEL mechanisms is that they are
always budget balanced in the following scenarios.

\begin{itemize}
        \item $[\theta]_1=U$ and $k=0$

        \item $[\theta]_{k+1}=[\theta]_k$ and $k \in \{1, \LL,  n-1\}$

        \item $[\theta]_n=L$ and $k=n$
\end{itemize}

Using this property, we will prove that the OEL mechanisms are the only collectively
undominated non-deficit Groves mechanisms that are anonymous and linear.

We first show that the OEL mechanisms are collectively undominated. 

\begin{theorem} 
\label{thm:oel}
For multi-unit auctions with unit demand, there is no non-deficit Groves mechanism that collectively dominates an OEL
mechanism.
\end{theorem}

By using Lemma~\ref{lem:anon}, we only need to prove this for the case of
anonymous Groves mechanisms.

\begin{lemma} \label{lem:oel} For multi-unit auctions with unit demand, there is no non-deficit anonymous Groves
mechanism that collectively dominates an OEL mechanism.  \end{lemma}

\begin{Proof}
We first prove: {\em no OEL mechanism with index $k\in \{1,\LL,n-1\}$ is
collectively dominated by a non-deficit anonymous Groves mechanism.}

Suppose a non-deficit anonymous Groves mechanism $r$
collectively dominates an OEL mechanism with index $k\in \{1,\LL,n-1\}$.  We
use $r^{OEL}$ to denote this OEL mechanism.
For any $i$ and $\theta_{-i}$, we define the following function:
\[
\Delta(\theta_{-i})=r(\theta_{-i})-r^{OEL}(\theta_{-i}).
\]

Since $r$ collectively dominates $r^{OEL}$, we have that for any $\theta$,
$\sum_{i=1}^n\Delta(\theta_{-i})\ge 0$.  

We also have that, whenever $[\theta]_{k+1}=[\theta]_{k}$, the OEL mechanism is
budget balanced. That is, under $r^{OEL}$, the agents' total payment is $0$; in
this case, since $r$ is non-deficit, we must have
$\sum_{i=1}^n\Delta(\theta_{-i})=0$.

Now we claim that $\Delta(\theta_{-i})=0$ for all $\theta_{-i}$.
Let $C(\theta_{-i})$ be the number of bids among $\theta_{-i}$ that equal
$[\theta_{-i}]_{k}$.  Hence, we must show that for all $\theta_{-i}$ with
$C(\theta_{-i})\ge 1$, we have $\Delta(\theta_{-i})=0$.

We now prove it by induction on the value of $C(\theta_{-i})$ (backwards, from $n-1$ to
$1$).

\II

\noindent \emph{Base case: $C(\theta_{-i})=n-1$.} 

Suppose there is a $\theta_{-i}$ with $C(\theta_{-i}) = n-1$.  That is,
all the bids in $\theta_{-i}$ are identical.  When $\theta_i$ is also equal
to the bids in $\theta_{-i}$, all bids in $\theta$ are the same so that
$[\theta]_{k+1}=[\theta]_{k}$. Hence, by our earlier observation, we have
$\sum_{j=1}^n\Delta(\theta_{-j})=0$.  But we know that for all $j$,
$\theta_{-j}$ is the same set of bids.  Hence $\Delta(\theta_{-i})=0$ for
all $\theta_{-i}$ when $C(\theta_{-i})= n-1$.  \II

\noindent \emph{Induction step.} 

Let us assume that for all $\theta_{-i}$, if $C(\theta_{-i})\ge p$ (where
$p \in \{2, \LL, n-1\}$), then $\Delta(\theta_{-i})=0$.  Now we consider
any $\theta_{-i}$ with $C(\theta_{-i}) = p-1$. When $\theta_i$ is equal to
$[\theta_{-i}]_{k}$, we have $[\theta]_{k}=[\theta]_{k+1}$, which implies
that $\sum_{j=1}^n\Delta(\theta_{-j})=0$.  For all $j$ with
$\theta_j=[\theta_{-i}]_{k}$, $\Delta(\theta_{-j})=\Delta(\theta_{-i})$, and
for other $j$, $C(\theta_{-j}) = p$.  Therefore, by the induction
assumption, $\sum_{j=1}^n \Delta(\theta_{-j})$ is a positive multiple of
$\Delta(\theta_{-i})$, which implies that $\Delta(\theta_{-i})=0$.

By induction, we have shown that $\Delta(\theta_{-i})=0$ for all
$\theta_{-i}$.  This implies that $r$ and $r^{OEL}$ are identical. Hence,
no other non-deficit anonymous Groves mechanism collectively dominates an OEL
mechanism with index $k\in \{1,\LL,n-1\}$.
\\
\\
\indent Now we prove: {\em the OEL mechanism with index $k=0$ is not
collectively dominated by a different non-deficit anonymous Groves mechanism.}

Suppose a non-deficit anonymous Groves mechanism $r$
collectively dominates an OEL mechanism with index $k=0$.
We use $r^{OEL}$ to denote this OEL mechanism.
For any $i$ and $\theta_{-i}$, we define the following function:
\[
\Delta(\theta_{-i})=r(\theta_{-i})-r^{OEL}(\theta_{-i}).
\]

Since $r$ collectively dominates $r^{OEL}$, we have that for any $\theta$,
$\sum_{i=1}^n\Delta(\theta_{-i}) \ge 0$.  
We also have that, whenever $[\theta]_1=U$, under $r^{OEL}$, the agents' total payment is $0$;
in this case, because $r$ is non-deficit,
we must have $\sum_{i=1}^n\Delta(\theta_{-i})=0$.

Now we claim that $\Delta(\theta_{-i})=0$ for all $\theta_{-i}$.
Let $C(\theta_{-i})$ be the number of bids among $\theta_{-i}$ that equal
$U$.  Hence, we must show that for all $\theta_{-i}$ with
$C(\theta_{-i})\ge 0$, we have $\Delta(\theta_{-i})=0$.

We now prove it by induction on the value of $C(\theta_{-i})$ (backwards, from $n-1$ to
$0$).

\II

\noindent \emph{Base case: $C(\theta_{-i}=n-1$.} 

Suppose there is a $\theta_{-i}$ with $C(\theta_{-i}) = n-1$.  That is,
all the bids in $\theta_{-i}$ are equal to $U$.  When $\theta_i$ is also equal
to the bids in $U$, by our earlier observation, we have
$\sum_{j=1}^n\Delta(\theta_{-j})=0$.  But we know that for all $j$,
$\Delta(\theta_{-j})$ is the same value.  Hence $\Delta(\theta_{-i})=0$ for
all $\theta_{-i}$ when $C(\theta_{-i})= n-1$.  \II

\noindent \emph{Induction step.} 

Let us assume that for all $\theta_{-i}$, if $C(\theta_{-i})\ge p$ (where
$p \in \{2, \LL, n-1\}$), then $\Delta(\theta_{-i})=0$.  Now we consider
any $\theta_{-i}$ with $C(\theta_{-i}) = p-1$. When $\theta_i$ is equal to
$U$, we have $[\theta]_1=U$, which implies
that $\sum_{j=1}^n\Delta(\theta_{-j})=0$.  For all $j$ with
$\theta_j=U$, $\Delta(\theta_{-j})=\Delta(\theta_{-i})$, and
for other $j$, $C(\theta_{-j}) = p$.  Therefore, by the induction
assumption, $\sum_{j=1}^n \Delta(\theta_{-j})$ is a positive multiple of
$\Delta(\theta_{-i})$, which implies that $\Delta(\theta_{-i})=0$.

By induction, we have shown that $\Delta(\theta_{-i})=0$ for all
$\theta_{-i}$.  This implies that $r$ and $r^{OEL}$ are identical. Hence,
no other non-deficit anonymous Groves mechanism collectively dominates the OEL
mechanism with index $k=0$.
\\
\\
\indent It remains to prove: {\em the OEL mechanism with index $k=n$ is
not collectively dominated by a different non-deficit anonymous Groves mechanism.}

This case is similar to the case of $k=0$ and we omit it here.
\end{Proof}

We now proceed to show that within the family of anonymous and linear non-deficit Groves
mechanisms, the OEL mechanisms are the only ones that are collectively
undominated.  Actually, they are also the only ones that are individually
undominated, which is a stronger claim since being individually undominated is
a weaker property.

\begin{theorem} \label{thm:characterize} For multi-unit auctions with unit demand, if an anonymous linear non-deficit
Groves mechanism is individually undominated, then it must be an OEL mechanism.
\end{theorem}

Before proving this theorem, let us introduce the following lemma.

\begin{lemma} 
\label{lm:measure}
Let $I$ be the set of points $(s_1,s_2,\ldots,s_k)$ ($U\ge s_1\ge s_2\ge\ldots\ge s_k\ge L$) that satisfy $Q_0+Q_1s_1+Q_2s_2+\ldots+Q_ks_k=0$ (the $Q_i$ are constants). If the measure of $I$ is positive (Lebesgue measure on $R^k$), then $Q_i=0$ for all $i$.
\end{lemma}

\begin{Proof} If $Q_i\neq 0$ for some $i$, then
  for any $U\ge s_1\ge s_2\ge\ldots\ge s_{i-1} \ge s_{i+1}\ge \ldots\ge
  s_k\ge L$, to make $Q_0+Q_1s_1+Q_2s_2+\ldots+Q_ks_k=0$, $s_i$ can take at
  most one value. As a result the measure of $I$ must be $0$.
\end{Proof}

Now we are ready to prove Theorem~\ref{thm:characterize}.

\begin{Proof}
Let $r$ be a non-deficit
anonymous linear Groves mechanism.  We recall that a Groves mechanism is
anonymous and linear if $r$ is a linear function defined as
$r(\theta_{-i})=a_0+\sum\limits_{j=1}^{n-1}a_j[\theta_{-i}]_j$
(where $[\theta_{-i}]_j$ is the $j$th highest type among $\theta_{-i}$, and the $a_j$'s
are constants).

Under multi-unit auctions with unit demand, the total VCG payment equals $m[\theta]_{m+1}$ ($m$ times the $(m+1)$th bid).
Under $r$, the agents' total payment 
equals
\[m[\theta]_{m+1}-\sum_{i=1}^nr(\theta_{-i})
=m[\theta]_{m+1}-na_0-\sum\limits_{i=1}^n\sum\limits_{j=1}^{n-1}a_j[\theta_{-i}]_j.\]
The above total payment is a linear function in terms of the types among
$\theta$. For simplicity, we rewrite the total payment as
$C_0+C_1[\theta]_1+C_2[\theta]_2+\ldots+C_n[\theta]_n$. The $C_i$ are constants
determined by the $a_i$. We have 
\begin{align*}
C_0&=-na_0\\
C_1&=-(n-1)a_1\\
C_2&=-a_1-(n-2)a_2\\
C_3&=-2a_2-(n-3)a_3\\
&\vdots\\
C_m&=-(m-1)a_{m-1}-(n-m)a_m\\
C_{m+1}&=-ma_{m}-(n-m-1)a_{m+1}+m\\
C_{m+2}&=-(m+1)a_{m+1}-(n-m-2)a_{m+2}\\
&\vdots\\
C_{n-1}&=-(n-2)a_{n-2}-a_{n-1}\\
C_n&=-(n-1)a_{n-1}
\end{align*}

Given any $\theta_{-i}$, for any possible value of $\theta_i$, we must have
$\sum\limits_{i=1}^n t_i(\theta) \ge 0$ (non-deficit).  That is, for any
$\theta_{-i}$, we have $\inf\limits_{\theta_i} \sum\limits_{i=1}^n t_i(\theta)
\ge 0$. If for some $\theta_{-i}$, we have $\inf\limits_{\theta_i}
\sum\limits_{i=1}^n t_i(\theta) > \epsilon$ ($\epsilon>0$), then we can reduce
the payment of agent $i$ by $\epsilon$ without violating the non-deficit
constraint, when the other agents' types are $\theta_{-i}$. Therefore, if the
mechanism is individually undominated, then for any $\theta_{-i}$, we have
$\inf\limits_{\theta_i} \sum\limits_{i=1}^n t_i(\theta)=0$. 

We denote $[\theta_{-i}]_j$ by $s_j$ ($j=1,\ldots,n-1$).
That is, $s_1\ge s_2\ge \ldots\ge s_{n-1}$.

The expression $\inf\limits_{\theta_i}\sum\limits_{i=1}^n t_i(\theta)$ then equals the minimum of the following expressions:

\[\inf\limits_{L\le\theta_i\le s_{n-1}}\sum\limits_{i=1}^n t_i(\theta)\]

\[\inf\limits_{s_{n-1}\le\theta_i\le s_{n-2}}\sum\limits_{i=1}^n t_i(\theta)\]

\[\vdots\]

\[\inf\limits_{s_{2}\le\theta_i\le s_{1}}\sum\limits_{i=1}^n t_i(\theta)\]

\[\inf\limits_{s_1\le\theta_i\le U}\sum\limits_{i=1}^n t_i(\theta)\]

We take a closer look at $\inf\limits_{L\le\theta_i\le
s_{n-1}}\sum\limits_{i=1}^n t_i(\theta)$.  When $L\le\theta_i\le s_{n-1}$, the
$j$th highest type $[\theta]_j=s_j$ for $j=1,\ldots,n-1$, and the $n$th highest
type $[\theta]_n=\theta_i$ (this case corresponds to agent $i$ being the agent
with the lowest type). 
We have \[\inf\limits_{L\le\theta_i\le
s_{n-1}}\sum\limits_{i=1}^n t_i(\theta)
=\inf\limits_{L\le\theta_i\le s_{n-1}}(C_0+C_1s_1+C_2s_2+\ldots+C_{n-1}s_{n-1}+C_n\theta_i)\]
\[=\min\{C_0+C_1s_1+\ldots+C_{n-1}s_{n-1}+C_nL, 
C_0+C_1s_1+\ldots+C_{n-1}s_{n-1}+C_ns_{n-1}\}.\] 
That is, because the expression is linear, the minimum is reached when $\theta_i$ is set to either the lower bound $L$ or the upper bound $s_{n-1}$.

Similarly, we have 
\[\inf\limits_{s_{n-1}\le\theta_i\le
s_{n-2}}\sum\limits_{i=1}^n t_i(\theta)
=\min\{C_0+C_1s_1+\ldots+C_{n-2}s_{n-2}+C_{n-1}s_{n-1}+C_ns_{n-1},\]
\[C_0+C_1s_1+C_2s_2+\ldots+C_{n-2}s_{n-2}+C_{n-1}s_{n-2}+C_ns_{n-1}\},\]
\[\vdots\]
\[\inf\limits_{s_{2}\le\theta_i\le s_{1}}\sum\limits_{i=1}^n t_i(\theta)
=\min\{C_0+C_1s_1+C_2s_1+C_3s_2+\ldots+C_ns_{n-1},\]
\[C_0+C_1s_1+C_2s_2+C_3s_2+\ldots+C_ns_{n-1}\},\]
\[\inf\limits_{s_1\le\theta_i\le U}\sum\limits_{i=1}^n t_i(\theta)
=\min\{C_0+C_1U+C_2s_1+\ldots+C_ns_{n-1},\]
\[C_0+C_1s_1+C_2s_1+\ldots+C_ns_{n-1}\}.\]

Putting all the above together, we have that for any $U\ge s_1\ge s_2\ge\ldots\ge s_{n-1}\ge L$, the minimum of the following expressions is $0$.

\begin{itemize}
\item $(n)$: $C_0+C_1s_1+C_2s_2+\ldots+C_{n-1}s_{n-1}+C_nL$ 

\item $(n-1)$: $C_0+C_1s_1+C_2s_2+\ldots+C_{n-1}s_{n-1}+C_ns_{n-1}$

\item $(n-2)$: $C_0+C_1s_1+C_2s_2+\ldots+C_{n-2}s_{n-2}+C_{n-1}s_{n-2}+C_ns_{n-1}$

\item $\vdots$

\item $(2)$: $C_0+C_1s_1+C_2s_2+C_3s_2+\ldots+C_ns_{n-1}$

\item $(1)$: $C_0+C_1s_1+C_2s_1+C_3s_2+\ldots+C_ns_{n-1}$

\item $(0)$: $C_0+C_1U+C_2s_1+C_3s_2+\ldots+C_ns_{n-1}$ 

\end{itemize}

The above expressions are numbered from $0$ to $n$.  Let $I(i)$ be the set of
points $(s_1,s_2,\ldots,s_{n-1})$ ($U\ge s_1\ge s_2\ge\ldots\ge s_{n-1}\ge L$)
that make expression $(i)$ equal to $0$. There must exist at least one $i$ such
that the measure of $I(i)$ is positive. According to Lemma~\ref{lm:measure},
expression $(i)$ must be the constant $0$.  

If expression $(0)$ is constant $0$, then the total payment under $r$
is $0$ whenever the highest type is equal to the upper bound $U$.
That is, for any $\theta$, the total payment 
$C_0+C_1[\theta]_1+C_2[\theta]_2+\ldots+C_n[\theta]_n$ must be a constant
multiple of $U-[\theta]_1$ (the total payment is a linear function).  We have
$C_0=-UC_1$ and $C_j=0$ for $j\ge 2$. It turns out that the above equalities of
the $C_j$ completely determine the values of the $a_j$ (the values of the $a_j$ can be solved for based on the $C_j$ by pure algebraic manipulations), and the corresponding
mechanism is the OEL mechanism with index $k=0$.
If expression $(i)$ is constant for other values of $i$, then the 
corresponding mechanism is the OEL mechanism with another index.
\end{Proof}

Hence, we have the following complete characterization in this context.

\begin{corollary} For multi-unit auctions with unit demand, a non-deficit anonymous linear Groves mechanism is
individually / collectively undominated if and only if it is an OEL mechanism.
\end{corollary}

\begin{Proof} This corollary can be proved by combining Theorem~\ref{thm:oel} and Theorem~\ref{thm:characterize}, as well as the fact that a collectively undominated
mechanism is also individually undominated. 
\end{Proof}

The above corollary also shows that if we consider only Groves mechanisms that
are non-deficit, anonymous, and linear in the setting of multi-unit auctions
with unit demand, then individual undominance and collective undominance are
equivalent.  Thus, we have characterized all individually/collectively
undominated Groves mechanisms that are non-deficit, anonymous, and linear for
multi-unit auctions with unit demand.

\section{The public project problem}
\label{sec:pp-equal}

We now study a well known class of decision problems, namely public project problems~\cite<see, {\em e.g.},>{Mas-Colell95,Mou88,Moo06}.
In this setting a set of $n$ agents needs to decide on financing a project of cost $c$. An agent's type
is her private valuation for the project if it takes place.
We consider two versions of the problem.

\subsection{Equal participation costs}

In this case if the project takes place, each agent contributes the
same share, $c/n$, so as to cover the total cost.  Hence the
participation costs of all agents are the same.
So the problem is defined as follows.

\paragraph{Public project problem} \label{exa:public1}
\mbox{} \\
\NI
Consider
$
(D, \Theta_1, \LL, \Theta_n, v_1, \LL, v_n),
$
where 

\begin{itemize}
\item[$\bullet$] $D = \{0, 1\}$
(reflecting whether a project is canceled or takes place),

\item[$\bullet$] for all $i \in \{1, \LL, n\}$,
$\Theta_i = [0,c]$, where $c > 0$,

\item[$\bullet$] for all $i \in \{1, \LL, n\}$, $v_i(d, \theta_i) := d (\theta_i - \frac{c}{n})$,

\eat{
\item[$\bullet$] $
        f(\theta) :=
        \left\{
        \begin{array}{l@{\extracolsep{3mm}}l}
        1    & \mathrm{if}\  \sum_{i = 1}^{n} \theta_i \geq c \\
        0       & \mathrm{otherwise}
        \end{array}
        \right.
$
}
\end{itemize}

When the agents employ a payment-based mechanism to decide on the project, then in addition to $c/n$, each agent also has to
pay or receive the payment, $t_i(\theta)$, imposed by the mechanism.
By the result of Holmstrom~\cite{Holmstrom79:Groves}, the only efficient and strategy-proof payment-based mechanisms in this domain are Groves mechanisms.
To determine the efficient outcome for a given type vector $\theta$, note that $\sum_{i =
  1}^{n} v_i(d, \theta_i) = d(\sum_{i = 1}^{n} \theta_i - c)$. Hence efficiency here for a mechanism $(f,t)$ means that $f(\theta) = 1$ if $\sum_{i = 1}^{n} \theta_i \geq c$ and $f(\theta) = 0$ 
otherwise, {\em i.e.}, the project takes place if and only if the declared total
value that the agents have for the project exceeds its cost.

We first observe the following result.
\begin{proposition}
\label{prop:BC-VCG}
In the public project problem with equal participation costs, the BC mechanism coincides with VCG.
\end{proposition}

\begin{Proof}
It suffices to check that in equation (\ref{eq:BC}) it holds that  $S_i^{BCGC}$ $(\theta_{-i}) = 0$ for all $i$ and all $\theta_{-i}$. 
Since VCG is a non-deficit mechanism, we have 
 $S_i^{BCGC} (\theta_{-i}) \geq 0$, as the term $S_i^{BCGC} (\theta_{-i})$ is a sum of payments for some type vector. 
 Hence all we need is to show that there is a value for $\theta_i'$ that makes 
the expression in  (\ref{eq:BC}) equal to $0$. Checking this is quite simple.
If $\sum_{j \neq i} \theta_j < \frac{n-1}{n} c$, 
then we take $\theta_i' := 0$ and otherwise
$\theta_i' := c$. In the former case the efficient outcome is to not implement the project whereas in the latter case, the opposite occurs. 
It is easy to check that in both cases we have $S_i^{BCGC} (\theta_{-i}) = 0$.
\end{Proof}

We now show that in fact VCG cannot be improved upon.
Before stating our result, we would like to note that one ideally would like to have a mechanism that is budget-balanced, {\em i.e.}, $\sum_i t_i(\theta) = 0$ for all $\theta$, so that in total the agents only pay the cost of the project and no more. However this is not possible and as explained in~\cite<page 861-862>{Mas-Colell95}, for the public
project problem no mechanism exists that is efficient,
strategy-proof and budget balanced. 
Our theorem below considerably strengthens this result, showing that VCG is optimal with respect to minimizing the total payment of the agents.

\begin{theorem}  
\label{thm:pp-collect-undom}
  In the public project problem there exists no non-deficit Groves mechanism
  that collectively dominates the VCG mechanism.
\end{theorem}

As with the case of unit-demand auctions, we first establish the desired conclusion for anonymous Groves mechanisms and then extend it to arbitrary ones by Lemma~\ref{lem:anon}. Notice that VCG is anonymous in this setting and hence we can apply Lemma~\ref{lem:anon}{({\em ii})}.

\begin{lemma}
\label{lem:pp-anonymous}
In the public project problem there exists no anonymous non-deficit Groves
mechanism that collectively dominates the VCG mechanism.
\end{lemma}

\begin{Proof}
Suppose that an anonymous non-deficit Groves mechanism $(r_1,...,r_n)$ 
 exists that collectively dominates VCG. By anonymity, for all $i \in
\{1, \LL, n\}$ ~$r_i = r$, for some function
 $r :
[0,c]^{n-1} \rightarrow \mathbb R$.
Hence 
\begin{equation}
\label{eq:assumption}
\forall ~\theta \in [0,c]^{n} ~~
\sum_{i = 1}^{n} r(\theta_{-i}) \geq 0
\end{equation}
We will show that then for all
$x\in[0,c]^{n-1}$, $r(x) = 0$ and thus $r$ coincides with VCG.

We divide our proof into two cases.

\noindent {\it Case 1:} The vector $x$ satisfies $\displaystyle\sum_{i=1}^{n-1} x_i \geq \frac{n-1}{n} c$.

Given such an $x$, define $C(x) = |\{i : x_i = c\}|$, {\em i.e.}, given a
vector $x$ of $n-1$ types, $C(x)$ is the number of agents who submitted
$c$.  Define the following predicate:

$$P(k): \ \fa x\in[0,c]^{n-1}~ ((C(x) = k ~\wedge ~\sum_{i=1}^{n-1} x_i \geq \frac{n-1}{n} c ) \myra r(x) = 0)$$

We now prove that $P(k)$ holds for all $k\in\{0,\LL,n-1\}$, using induction
(going backwards from $n-1$).
Let $t_i(\theta) = VCG_i(\theta) - r(\theta_{-i})$ be the payment function of agent $i$ under the mechanism $r$. 
\\
\\
\noindent \emph{Base case.} 

Let $x$ be such that $C(x) = n-1$. Consider $\theta := (c,\LL,c) \in
[0,c]^n$. Then for all $i \in \{1, \LL, n\}$, $\theta_{-i} = x$. Clearly
$f(\theta) = 1$ and no agent is paying anything under the VCG mechanism in this instance, {\em i.e.}, 
$VCG_i(\theta) = 0$.

Since $r$ is a non-deficit mechanism
$$
0 \leq \sum_{i = 1}^{n} t_i(\theta) = \sum_{i = 1}^{n} VCG_i(\theta) - \sum_{i = 1}^{n} r(\theta_{-i}) = -\sum_{i = 1}^{n} r(\theta_{-i}) = -nr(x),
$$

But then by (\ref{eq:assumption}) we have $r(x)=0$.

\II

\noindent \emph{Induction step.} 

Assume $P(k)$ holds for some $k\geq 1$. We will prove $P(k-1)$. Let $x$ be
such that $C(x) = k-1$ (note that $x$ may have zero $c$'s). Since $r$ is permutation independent, 
we can assume without loss of generality that the elements
of $x$ are sorted in descending order ({\em i.e.}, $r(x)$ does not change by such a reordering). 
Consider the type vector $\theta =
(c, x)$, that is the concatenation of $(c)$ and $x$.  Hence $\theta$ starts
with $k$ $c$'s and the rest is like the rest of $x$.  Note that for $i \in
\{1,\LL,k\}$, $\theta_{-i} = x$ and $C(\theta_{-i}) = k-1$.  For $i \in
\{k+1,\LL,n\}$, $C(\theta_{-i}) = k$, therefore by induction hypothesis,
$r(\theta_{-i}) = 0$. This means that $\sum_{i = 1}^{n} r(\theta_{-i}) = kr(x)$.

Furthermore, $f(\theta) =1$ since $\theta$ has at least one $c$, and no agent is paying payment under the VCG mechanism. To see this, if $k\geq 2$, then for every agent under $\theta$, there is another agent who submitted $c$ hence the agent is not pivotal. If $k=1$, then no agent can alter the decision outcome by the fact that $\sum x_i \geq \frac{n-1}{n} \cdot c$, hence no agent is pivotal in this case as well. 
Thus, for all $i \in \{1, \LL, n\}$,
$VCG_i(\theta) = 0$, and because $r$ is non-deficit  
$$0\leq \sum_{i =1}^{n} t_i(\theta) = - \sum_{i =1}^{n} r(\theta_{-i}) = -kr(x) $$ 
But then by (\ref{eq:assumption}) we have that $r(x) = 0$. This concludes the induction step and consequently $r(x)=0$ for all vectors $x$ that belong to Case 1.
\II

\noindent {\it Case 2:} The vector $x$ satisfies $\displaystyle\sum_{i=0}^{n-1} x_i < \frac{n-1}{n} c$. The proof for this case uses a completely symmetric argument to that of Case $1$. We include it below for the sake of completeness.

Define $C'(x) = |\{i : x_i = 0\}|$. In analogy to the predicate $P(k)$ of Case $1$, we define the following predicate:

$$P'(k): \ \fa x\in[0,c]^{n-1} ((C'(x) = k \wedge \sum_{i=0}^{n-1} x_i < \frac{n-1}{n} c) \myra r(x)=0)$$

We now prove that $P'(k)$ holds for all $k\in\{0,...,n-1\}$, using induction
(going backwards from $n-1$).
\\
\\
\noindent \emph{Base case.} 

Let $x$ be such that $C'(x) = n-1$, {\em i.e.}, the zero vector. Consider $\theta := (0,...,0) \in
[0,c]^n$. Then for all $i \in \{1, \LL, n\}$, $\theta_{-i} = x$. Clearly
$f(\theta) = 0$ and no agent is paying anything under the VCG mechanism. Hence if $t_i(\theta)$ is the payment paid by agent $i$, then $\sum t_i(\theta) = -\sum r(\theta_{-i}) = -n r(x)$.

Since $r$ is a non-deficit mechanism, 
$$
0 \leq \sum_{i = 1}^{n} t_i(\theta) = -nr(x)
$$
Then by (\ref{eq:assumption}) this implies that $r(x)=0$.

\II

\noindent \emph{Induction step.} 

Suppose $P'(k)$ holds for some $k\geq 1$. We will prove $P'(k-1)$. Let $x$ be
such that $C'(x) = k-1$. Since $r$ is permutation
independent, we can assume without loss of generality that the elements
of $x$ are sorted in increasing order so that all $0$'s are on the left side of $x$ (note that it may also be that $x$ does not have any $0$'s, since $k-1$ maybe equal to $0$). 
Consider the type vector $\theta =
(0, x)$. So $\theta$ starts
with $k$ $0$'s and the rest is like the rest of $x$.
Note that for $i \in
\{1,\LL,k\}$, $\theta_{-i} = x$ and $C'(\theta_{-i}) = k-1$.  For $i \in
\{k+1,\LL,n\}$, $C'(\theta_{-i}) = k$ and by induction hypothesis,
$r(\theta_{-i}) = 0$ and hence $\sum r(\theta_{-i}) = k r(x)$.

We note that $f(\theta) =0$ and that also no agent is paying payment under the VCG mechanism. To see this, it is enough to verify that no agent is pivotal, which follows by the fact that we are in the case that $\sum_{i=0}^{n-1} x_i < \frac{n-1}{n} c$. Since $\theta=(0,x)$, no agent can be pivotal.
Therefore $VCG_i(\theta) = 0$ for every $i \in \{1,\LL,n\}$. Since $r$ is non-deficit we have $0\leq -\sum r(\theta_{-i}) = -k r(x)$. By (\ref{eq:assumption}) we have $r(x)=0$.

This completes the proof of the induction step and hence Case $2$. Since Cases $1$ and $2$ cover all vectors $x\in [0, c]^{n-1}$, the proof of the Lemma is complete.
\end{Proof}

By using now Lemma \ref{lem:pp-anonymous} and Lemma~\ref{lem:anon}{({\em ii})}, the proof of Theorem \ref{thm:pp-collect-undom} is complete.

An interesting open question is whether other mechanisms that share some of the properties of the VCG mechanism are also collectively undominated. In particular, we have exhibited that VCG is a pay-only and anonymous mechanism. Are there other anonymous or pay-only mechanisms that are collectively undominated for the public project problem with equal participation costs?

We start with pay-only mechanisms.  We provide a general observation
that holds in many domains other than public project problems, showing
that the VCG mechanism dominates all other pay-only mechanisms.

\begin{lemma}
\label{lem:all-other}
Let $\textbf{r}$ be a Groves mechanism. Suppose that the following 
condition~\footnote{This is a slight generalization of the
 Potential for Universal Relevance Nullification (PURN) condition
introduced in \cite{Cavallo06:Optimal}. An agent satisfies PURN if
he can make his payment {\em under the VCG mechanism} equal to $0$ for any type vector $\theta_{-i}$ of the other agents. Here, the only difference is that
we consider all Groves mechanisms instead of just VCG.}
holds for
all $i \in \{1, \LL, n\}$:

$$ \forall \theta_{-i} \in \Theta_{-i} ~\exists b_i^* \in \Theta_i \mbox{ such that } VCG_i(b_i^*, \theta_{-i}) - r_i(\theta_{-i}) =0.$$
Then $\textbf{r}$ individually dominates all other pay-only Groves mechanisms.
\end{lemma}
The condition essentially says that every agent is always able to make his payment equal to $0$ for any type vector $\theta_{-i}$ of the other agents.

\begin{Proof}
Suppose that there exists a pay-only mechanism
$\textbf{r}' = (r_1',\LL,r_n')$ different from $\textbf{r} = (r_1,\LL,r_n)$ and not dominated by $\textbf{r}$. Then, for some
  $\theta \in \Theta$ and $i\in \{1, \LL n\}$, $r_{i}'(\theta_{-i}) > r_{i}(\theta_{-i})$. 
Let $b_{i}^*$ be the type of agent $i$ that satisfies the condition of the theorem. 
Consider $\theta' = (b_{i}^*, \theta_{-i})$. 
Then $VCG_{i}(\theta') = r_i(\theta_{-i})$. 

But then the payment of agent $i$ under mechanism $\textbf{r}'$ for the profile $\theta'$ is
\[
t_{i}'(\theta') = VCG_i(\theta') - r_i'(\theta_{-i}) < VCG_i(\theta') - r_i(\theta_{-i}) = 0,
\]
which is a contradiction, because $\textbf{r}'$ is a pay-only mechanism.
\end{Proof}

\begin{theorem}
\label{thm:pay-only}
Consider the public project problem with equal participation costs. Then for a pay-only Groves mechanism $\textbf{r}$, the following are equivalent:
\begin{enumerate}
\item $\textbf{r}$ is individually undominated,
\item $\textbf{r}$ is the VCG mechanism,
\item $\textbf{r}$ is collectively undominated.
\end{enumerate}
\end{theorem}

\begin{Proof}
\NI
$1 \rightarrow 2$. 
Consider a pay-only and individually undominated Groves mechanism $\textbf{r}$.
We claim that $\textbf{r}$ is the VCG mechanism. 

In the considered domain every agent $i$, given $\theta_{-i}$, can
force his VCG payment to be $0$ by declaring $b_i^* = c/n$. Indeed, we then would have
$VCG_i(c/n, \theta_{-i}) = 0$. Hence by Lemma \ref{lem:all-other} the VCG mechanism
individually dominates all other pay-only mechanisms. This means that there can be no
other individually undominated mechanism than VCG.
\II

\NI
$2 \rightarrow 3$ holds by Theorem~\ref{thm:pp-collect-undom} and $3\rightarrow 1$ holds by the definition.
\end{Proof}

The above theorem shows that for the public project problem with equal
participation costs, VCG is the only pay-only Groves mechanism that is
individually/collectively undominated.  In Appendix~\ref{app:twoagents}, we
show a similar result for anonymous Groves mechanisms, but only for the case of
two agents. That is, if there are exactly two agents, VCG is the only anonymous Groves
mechanism that is individually/collectively undominated.  
Further, for $n\geq 3$, Herv\'e Moulin (private
communication) observed that for public project problems with equal
participation costs, the VCG mechanism is not the only non-deficit Groves
mechanism that is collectively undominated. 



\subsection{The general case}
\label{sec:public1}

The assumption that we have made so far in the public project problem that
each agent's cost share is the same may not always be realistic. Indeed, it
may be argued that `richer' agents (such as larger enterprises) should
contribute more.  Does it matter if we modify the formulation of the
problem appropriately? The answer is `yes'.  First, let us formalize this
version of the problem.  We assume now that each initial utility function is of the form
\[
v_i(d, \theta_i) := d (\theta_i - c_i),
\]
where for all $i \in \C{1, \LL, n}$, \ $c_i > 0$ and $\sum_{i = 1}^{n} c_i = c$.

In this setting, $c_i$ is the share of the project cost to be
financed by agent $i$. 
We call the resulting problem the
\bfe{general public project problem}. It is taken from \cite<page
518>{Moo06}.
For this problem we have only two results, both concerning the individual dominance relation.

\begin{theorem} \label{thm:opt}
In the general public project problem the VCG mechanism individually dominates all other pay-only Groves mechanisms.
\end{theorem}

\begin{Proof}
Note that for any $i$ and any $\theta_{-i}$, agent $i$ can force his VCG payment to be $0$ by declaring $c_i$, since $t_i(c_i, \theta_{-i}) = 0$.
By Lemma~\ref{lem:all-other} the proof is complete.
\end{Proof}

The above theorem cannot be
extended to non-deficit Groves mechanisms, as is illustrated by the following theorem. The theorem below also shows that if there is an individually undominated mechanism in this setting, it cannot be a pay-only mechanism.

\begin{theorem}
\label{thm:instanceexists}
For any $n\geq 3$, an instance of the general public project problem
with $n$ agents exists for which the BC mechanism individually dominates the VCG
mechanism.
\end{theorem}

\begin{Proof}
  \NI We will show this for $n=3$. For $n>3$, it is fairly simple to extend the proof. We omit the details.
The VCG mechanism is non-deficit,
  hence it suffices to show by Proposition~\ref{prop:bc}$(ii)$ that the VCG and BC
  mechanisms do not coincide, for some choice of $c, c_1,c_2,c_3$, with
  $c_1+c_2+c_3 = c$.

To this end we need to find $\theta_2$ and $\theta_3$ so that
$S_1^{BCGC}(\theta_2, \theta_3) >0$.  Here
\[
S^{BCGC}_1(\theta_2, \theta_3) := \min_{\theta_1' \in \Theta_1}((R_1 + R_2 + R_3)-L),
\]
where for $\theta' := (\theta'_1, \theta_2, \theta_3)$
\[
L := (n-1) \sum_{k=1}^n v_k(f(\theta'), \theta'_k),
\]
\[
R_1 = \max_{d \in D} \sum_{j \neq 1} v_j(d, \theta'_j) = \max \{0, \theta_2
+ \theta_3 -(c_2 + c_3)\},
\]
\[
R_2 = \max_{d \in D} \sum_{j \neq 2} v_j(d, \theta'_j) = \max \{0,
\theta'_1 + \theta_3 -(c_1 + c_3)\},
\]
\[
R_3 = \max_{d \in D} \sum_{j \neq 3} v_j(d, \theta'_j) = \max \{0, \theta'_1 + \theta_2 -(c_1 + c_2)\}.
\]

Now, take $c = 100, c_1 = 10, c_2 = 40,
c_3 = 50$ and $\theta_2 := 10$, $\theta_3 := 70$.  
Then $
R_1 + R_2 + R_3 = \theta'_1 + 10 + \max \{0, \theta'_1 - 40\}.
$
Two cases arise.  

\NI
\emph{Case 1} $f(\theta') = 0$.

Then $L = 0$, so $(R_1 + R_2 + R_3) - L  \geq 10$.

\NI
\emph{Case 2} $f(\theta') = 1$.

Then $L = 2(\theta'_1 + \theta_2 + \theta_3 - 100) = 2 \theta'_1 - 40$, so
\begin{eqnarray*}
(R_1 + R_2 + R_3) - L & = & 50 - \theta'_1 + \max \{0, \theta'_1 - 40\} \\
&\geq &(50 - \theta'_1) +  (\theta'_1 - 40) \geq 10.
\end{eqnarray*}

This proves that $S^{BCGC}_1(\theta_2, \theta_3) \geq  10$. By taking any $\theta'_1 \in [40,100]$ we see that
in fact $S^{BCGC}_1(\theta_2, \theta_3) = 10$. 
\end{Proof}

By virtue of Theorem \ref{thm:opt} the BC mechanism in the above proof
is not pay-only. 

\section{Conclusions and future work}
\label{sec:conclusions}

The family of Groves mechanisms, which includes the well-known VCG mechanism
(also known as the Clarke mechanism), is a family of efficient and
strategy-proof mechanisms.  Unfortunately, the Groves mechanisms are
generally not budget balanced. That is, under such mechanisms, payments may
flow into or out of the system of the agents, resulting in deficits or reduced
utilities for the agents.  To identify non-deficit Groves mechanisms that give
the agents the highest utilities, we introduced two general measures for
comparing mechanisms in prior-free settings. Specifically, we say that a
non-deficit Groves mechanism $M$ {\em individually dominates} another
non-deficit Groves mechanism $M'$ if for every type profile, every agent's
utility under $M$ is no less than that under $M'$, and this holds with strict
inequality for at least one type profile and one agent.  We say that a
non-deficit Groves mechanism $M$ {\em collectively dominates} another
non-deficit Groves mechanism $M'$ if for every type profile, the agents' total
utility (social welfare) under $M$ is no less than that under $M'$, and this
holds with strict inequality for at least one type profile.  The above
definitions induce two partial orders on non-deficit Groves mechanisms.  This
paper mainly focused on studying the maximal elements corresponding to these two partial
orders.
\\

A number of interesting open problems remain. Specifically,

\begin{itemize}

\item We provided in Section~\ref{subsec:distinct}
two examples showing that collective undominance is strictly stronger
than individual undominance.
One example involves a discrete type space, while the other example involves discontinuous redistribution functions. It remains to be seen whether the two definitions of undominance coincide when the type space is smoothly connected and the redistribution functions are continuous.

\item We know from \cite{Guo10:Optimal} that the OEL mechanisms are not the
only collectively undominated mechanisms in multi-unit auctions with unit
demand, because there exist prior distributions under which other mechanisms
achieve strictly higher expected social welfare. That is, for multi-unit
auctions with unit demand, there exist other unknown collectively undominated
mechanisms (based on nonlinear redistribution functions).  However, it remains
to be seen whether there also exist collectively undominated mechanisms (other
than VCG) for public project problems.
\item We proposed two techniques for generating individually undominated mechanisms.
Can we also derive techniques for generating collectively undominated mechanisms?
\end{itemize}

\subsection*{Acknowledgements}
The authors would like to thank all three reviewers for their useful comments. We also thank Herv\'e Moulin for valuable discussions.
This work has been supported by the project DIACODEM of the Dutch organization for scientific research (NWO), and by the project AGT of the research funding program THALIS (co-financed by the European Social Fund-ESF and Greek national funds).
We also thank the National Science Foundation and the Alfred P.~Sloan
Foundation for support under Awards IIS-0812113, IIS-0953756, and
CCF-1101659, and a Sloan Fellowship.

\bibliographystyle{theapa}
\bibliography{../../mg}

\begin{thebibliography}{}

\bibitem[\protect\BCAY{Apt, Conitzer, Guo,\ \BBA\ Markakis}{Apt
  et~al.}{2008}]{Apt08:Welfare}
Apt, K., Conitzer, V., Guo, M., \BBA\ Markakis, E. \BBOP2008\BBCP.
\newblock \BBOQ Welfare undominated {G}roves mechanisms\BBCQ\
\newblock In {\Bem Proceedings of the Fourth Workshop on Internet and Network
  Economics (WINE)}, \BPGS\ 426--437, Shanghai, China.

\bibitem[\protect\BCAY{Bailey}{Bailey}{1997}]{Bailey97:Demand}
Bailey, M.~J. \BBOP1997\BBCP.
\newblock \BBOQ The demand revealing process: to distribute the surplus\BBCQ\
\newblock {\Bem Public Choice}, {\Bem 91}, 107--126.

\bibitem[\protect\BCAY{Cavallo}{Cavallo}{2006}]{Cavallo06:Optimal}
Cavallo, R. \BBOP2006\BBCP.
\newblock \BBOQ Optimal decision-making with minimal waste: Strategyproof
  redistribution of {VCG} payments\BBCQ\
\newblock In {\Bem Proceedings of the International Conference on Autonomous
  Agents and Multi-Agent Systems (AAMAS)}, \BPGS\ 882--889, Hakodate, Japan.

\bibitem[\protect\BCAY{Clarke}{Clarke}{1971}]{Clarke71:Multipart}
Clarke, E.~H. \BBOP1971\BBCP.
\newblock \BBOQ Multipart pricing of public goods\BBCQ\
\newblock {\Bem Public Choice}, {\Bem 11}, 17--33.

\bibitem[\protect\BCAY{Cramton, Gibbons,\ \BBA\ Klemperer}{Cramton
  et~al.}{1987}]{Cramton87:Dissolving}
Cramton, P., Gibbons, R., \BBA\ Klemperer, P. \BBOP1987\BBCP.
\newblock \BBOQ Dissolving a partnership efficiently\BBCQ\
\newblock {\Bem Econometrica}, {\Bem 55\/}(3), 615--632.

\bibitem[\protect\BCAY{de~Clippel, Naroditskiy,\ \BBA\ Greenwald}{de~Clippel
  et~al.}{2009}]{Clippel09:Destroy}
de~Clippel, G., Naroditskiy, V., \BBA\ Greenwald, A. \BBOP2009\BBCP.
\newblock \BBOQ Destroy to save\BBCQ\
\newblock In {\Bem Proceedings of the ACM Conference on Electronic Commerce
  (EC)}, \BPGS\ 207--214, Stanford, CA, USA.

\bibitem[\protect\BCAY{Faltings}{Faltings}{2005}]{Faltings05:Budget}
Faltings, B. \BBOP2005\BBCP.
\newblock \BBOQ A budget-balanced, incentive-compatible scheme for social
  choice\BBCQ\
\newblock In {\Bem Agent-Mediated Electronic Commerce (AMEC), LNAI, 3435},
  \BPGS\ 30--43.

\bibitem[\protect\BCAY{Groves}{Groves}{1973}]{Groves73:Incentives}
Groves, T. \BBOP1973\BBCP.
\newblock \BBOQ Incentives in teams\BBCQ\
\newblock {\Bem Econometrica}, {\Bem 41}, 617--631.

\bibitem[\protect\BCAY{Gujar\ \BBA\ Narahari}{Gujar\ \BBA\
  Narahari}{2011}]{Gujar11:Redistribution}
Gujar, S.\BBACOMMA\  \BBA\ Narahari, Y. \BBOP2011\BBCP.
\newblock \BBOQ Redistribution mechanisms for assignment of heterogeneous
  objects\BBCQ\
\newblock {\Bem J. Artif. Intell. Res. (JAIR)}, {\Bem 41}, 131--154.

\bibitem[\protect\BCAY{Guo}{Guo}{2011}]{Guo11:VCG}
Guo, M. \BBOP2011\BBCP.
\newblock \BBOQ {VCG} redistribution with gross substitutes\BBCQ\
\newblock In {\Bem Proceedings of the National Conference on Artificial
  Intelligence (AAAI)}, San Francisco, CA, USA.

\bibitem[\protect\BCAY{Guo}{Guo}{2012}]{Guo12:Worst}
Guo, M. \BBOP2012\BBCP.
\newblock \BBOQ Worst-case optimal redistribution of {VCG} payments in
  heterogeneous-item auctions with unit demand\BBCQ\
\newblock In {\Bem Proceedings of the Eleventh International Joint Conference
  on Autonomous Agents and Multi-Agent Systems (AAMAS)}, Valencia, Spain.

\bibitem[\protect\BCAY{Guo\ \BBA\ Conitzer}{Guo\ \BBA\
  Conitzer}{2008a}]{Guo08:Better}
Guo, M.\BBACOMMA\  \BBA\ Conitzer, V. \BBOP2008a\BBCP.
\newblock \BBOQ Better redistribution with inefficient allocation in multi-unit
  auctions with unit demand\BBCQ\
\newblock In {\Bem Proceedings of the ACM Conference on Electronic Commerce
  (EC)}, \BPGS\ 210--219, Chicago, IL, USA.

\bibitem[\protect\BCAY{Guo\ \BBA\ Conitzer}{Guo\ \BBA\
  Conitzer}{2008b}]{Guo08:Undominated}
Guo, M.\BBACOMMA\  \BBA\ Conitzer, V. \BBOP2008b\BBCP.
\newblock \BBOQ Undominated {VCG} redistribution mechanisms\BBCQ\
\newblock In {\Bem Proceedings of the Seventh International Joint Conference on
  Autonomous Agents and Multi-Agent Systems (AAMAS)}, \BPGS\ 1039--1046,
  Estoril, Portugal.

\bibitem[\protect\BCAY{Guo\ \BBA\ Conitzer}{Guo\ \BBA\
  Conitzer}{2009}]{Guo07:Worst}
Guo, M.\BBACOMMA\  \BBA\ Conitzer, V. \BBOP2009\BBCP.
\newblock \BBOQ Worst-case optimal redistribution of {VCG} payments in
  multi-unit auctions\BBCQ\
\newblock {\Bem Games and Economic Behavior}, {\Bem 67\/}(1), 69--98.

\bibitem[\protect\BCAY{Guo\ \BBA\ Conitzer}{Guo\ \BBA\
  Conitzer}{2010}]{Guo10:Optimal}
Guo, M.\BBACOMMA\  \BBA\ Conitzer, V. \BBOP2010\BBCP.
\newblock \BBOQ Optimal-in-expectation redistribution mechanisms\BBCQ\
\newblock {\Bem Artificial Intelligence}, {\Bem 174\/}(5-6), 363--381.

\bibitem[\protect\BCAY{Guo, Naroditskiy, Conitzer, Greenwald,\ \BBA\
  Jennings}{Guo et~al.}{2011}]{Guo11:Budget}
Guo, M., Naroditskiy, V., Conitzer, V., Greenwald, A., \BBA\ Jennings, N.~R.
  \BBOP2011\BBCP.
\newblock \BBOQ Budget-balanced and nearly efficient randomized mechanisms:
  Public goods and beyond\BBCQ\
\newblock In {\Bem Proceedings of the Seventh Workshop on Internet and Network
  Economics (WINE)}, Singapore.

\bibitem[\protect\BCAY{Holmstr{\"{o}}m}{Holmstr{\"{o}}m}{1979}]{Holmstrom79:Groves}
Holmstr{\"{o}}m, B. \BBOP1979\BBCP.
\newblock \BBOQ Groves' scheme on restricted domains\BBCQ\
\newblock {\Bem Econometrica}, {\Bem 47\/}(5), 1137--1144.

\bibitem[\protect\BCAY{Laffont\ \BBA\ Maskin}{Laffont\ \BBA\
  Maskin}{1997}]{Laffont82:The}
Laffont, J.\BBACOMMA\  \BBA\ Maskin, E. \BBOP1997\BBCP.
\newblock {\Bem The theory of incentives: An overview, in: W. Hildenbrand, ed.,
  Advances in economics theory, Econometric Society Monograph in Quantitative
  Economics}.
\newblock Cambridge University Press.

\bibitem[\protect\BCAY{Mas-Colell, Whinston,\ \BBA\ Green}{Mas-Colell
  et~al.}{1995}]{Mas-Colell95}
Mas-Colell, A., Whinston, M., \BBA\ Green, J.~R. \BBOP1995\BBCP.
\newblock {\Bem Microeconomic Theory}.
\newblock Oxford University Press.

\bibitem[\protect\BCAY{Moore}{Moore}{2006}]{Moo06}
Moore, J. \BBOP2006\BBCP.
\newblock {\Bem General Equilibrium and Welfare Economics: An Introduction}.
\newblock Springer.

\bibitem[\protect\BCAY{Moulin}{Moulin}{1988}]{Mou88}
Moulin, H. \BBOP1988\BBCP.
\newblock {\Bem Axioms of Cooperative Decision Making}.
\newblock Cambridge University Press.

\bibitem[\protect\BCAY{Moulin}{Moulin}{1986}]{Moulin86:Characterization}
Moulin, H. \BBOP1986\BBCP.
\newblock \BBOQ Characterizations of the pivotal mechanism\BBCQ\
\newblock {\Bem Journal of Public Economics}, {\Bem 31\/}(1), 53--78.

\bibitem[\protect\BCAY{Moulin}{Moulin}{2009}]{Moulin09:Almost}
Moulin, H. \BBOP2009\BBCP.
\newblock \BBOQ Almost budget-balanced {VCG} mechanisms to assign multiple
  objects\BBCQ\
\newblock {\Bem Journal of Economic Theory}, {\Bem 144\/}(1), 96--119.

\bibitem[\protect\BCAY{Myerson\ \BBA\ Satterthwaite}{Myerson\ \BBA\
  Satterthwaite}{1983}]{Myerson83:Efficient}
Myerson, R.\BBACOMMA\  \BBA\ Satterthwaite, M. \BBOP1983\BBCP.
\newblock \BBOQ Efficient mechanisms for bilateral trading\BBCQ\
\newblock {\Bem Journal of Economic Theory}, {\Bem 28}, 265--281.

\bibitem[\protect\BCAY{Porter, Shoham,\ \BBA\ Tennenholtz}{Porter
  et~al.}{2004}]{Porter04:Fair}
Porter, R., Shoham, Y., \BBA\ Tennenholtz, M. \BBOP2004\BBCP.
\newblock \BBOQ Fair imposition\BBCQ\
\newblock {\Bem Journal of Economic Theory}, {\Bem 118}, 209--228.

\end{thebibliography}

\appendix

\section{Uniqueness of VCG for the case of two agents}
\label{app:twoagents}
\begin{theorem}
\label{thm:n=2}
Consider the public project problem with equal participation costs. When the number of agents is
$n=2$, then for a non-deficit, and anonymous Groves mechanism $\textbf{r}$, the
following are equivalent:
\begin{enumerate}
\item $\textbf{r}$ is individually undominated,
\item $\textbf{r}$ is the VCG mechanism,
\item $\textbf{r}$ is collectively undominated.
\end{enumerate}
\end{theorem}

\begin{Proof}
  As in the proof of Theorem~\ref{thm:pay-only} it suffices to show that
  $1\rightarrow 2$.  So take a non-deficit, anonymous, and individually undominated
  Groves mechanism, determined by the function $r$. 
  
  For $x\in [0, c]$, take $\theta := (x,x)$. If $x\geq c/2$, then the
  efficient outcome is $f(\theta)=1$ and no agent is pivotal, hence
  the total VCG payment is $0$. If $x< c/2$, then the project is not built
  and again no agent is pivotal. Hence in both cases the VCG payment is
  $0$. If $t$ is the payment function corresponding to $r$, then we have
  that $t_1(\theta) + t_2(\theta) = -2 r(x)$. Since $r$ is non-deficit,
  we have that for every $x\in [0, c]$, $r(x)\leq 0$. But since $r$ is
  individually undominated, it cannot be the case that $r(x) < 0$ for some $x$,
  because then the VCG mechanism would dominate $r$. Hence $r$ coincides with the
  VCG mechanism.
\end{Proof}


\end{document}